\documentclass[12pt, a4paper, table]{article} 
\usepackage[includeheadfoot,
            marginratio={1:1,2:3}, 
            width=452pt, 
            height=720pt,]{geometry}
\usepackage{physics}
\usepackage{amsmath}
\usepackage{amsfonts}
\usepackage{amssymb}
\usepackage{amsthm}
\usepackage{bbm,bm}
\usepackage{mathtools, comment}
\usepackage{slashed}
\usepackage{mathalfa}
\usepackage{graphicx,caption,subfigure}
\usepackage{multirow, multicol}
\usepackage{tikz} 
\usepackage{tikz-cd}
\usetikzlibrary{decorations.pathreplacing,calc,tikzmark}
\usepackage{booktabs}
\usepackage{adjustbox}
\usepackage[utf8]{inputenc}
 \usepackage[splitrule]{footmisc}

\usepackage{caption}
\usepackage{placeins}
\usepackage{empheq}
\usepackage{paralist}
\usepackage{cite}
\usepackage[normalem]{ulem}
\usepackage{soul}

\usepackage[colorlinks=false,urlbordercolor=red
]{hyperref}
\usepackage{tensor}

\newcommand{\nc}{\newcommand}
\nc{\lb}{\llbracket}
\nc{\rb}{\rrbracket}
\nc{\gl}{\llbracket}
\nc{\gr}{\rrbracket}
\nc{\del}{\partial}
\nc{\tri}{\hspace{-3.5pt}\vartriangle\hspace{-3.5pt}}
\nc{\blacktri}{\blacktriangle}

\definecolor{rossoCP3}{cmyk}{0,.88,.77,.40}

\nc{\eq}[1]{\begin{equation}
                     \begin{split} #1 \end{split}
                     \end{equation}}
\nc{\ov}{\overline}

\nc{\fa}{\hat}
\nc{\fb}{\MakeUppercase}
\nc{\fc}{\tilde }
\nc{\Lie}{{\cal L}} 
\nc{\lambdabar}{{\mkern0.75mu\mathchar '26\mkern -9.75mu\lambda}}

\allowdisplaybreaks[2]
\numberwithin{equation}{section}

\DeclareUnicodeCharacter{2212}{-}
\begin{document}

\vspace*{-1.5cm}
\begin{flushright}
  {\small
  LMU-ASC 35/23\\ MPP-2023-262
  }
\end{flushright}

\vspace{1.5cm}
\begin{center}
  {\Large \bf Shedding black hole light \\ on the emergent string conjecture} 
\vspace{0.35cm}

\end{center}

\vspace{0.35cm}
\begin{center}
{\large Ivano Basile$^a$,
Dieter L\"ust$^{a,b}$ and Carmine Montella$^{b}$
}
\end{center}

\vspace{0.1cm}
\begin{center} 
\emph{
$^a${\it Arnold Sommerfeld Center for Theoretical Physics,\\
Ludwig-Maximilians-Universit\"at M\"unchen, 80333 M\"unchen, Germany},   \\[0.1cm] 
 \vspace{0.3cm}
$^b$Max-Planck-Institut f\"ur Physik (Werner-Heisenberg-Institut), \\[.1cm] 
   F\"ohringer Ring 6,  80805 M\"unchen, Germany, 
   \\[0.1cm] 
    } 
\end{center} 

\vspace{0.5cm}

\begin{abstract}
    Asymptotically massless towers of species are ubiquitous in the string landscape when infinite-distance limits are approached. Due to the remarkable properties of string dualities, they always comprise Kaluza-Klein states or higher-spin excitations of weakly coupled, asymptotically tensionless critical strings. The connection between towers of light species and small black holes warrants seeking a bottom-up rationale for this dichotomoy, dubbed emergent string conjecture. In this paper we move a first step in this direction, exploring bottom-up constraints on towers of light species motivated purely from the consistency of the corresponding thermodynamic picture for small black holes. These constraints shed light on the allowed towers in quantum gravity, and, upon combining them with unitarity and causality constraints from perturbative graviton scattering, they provide evidence for the emergent string scenario with no reference to a specific ultraviolet completion.
\end{abstract}

\thispagestyle{empty}
\clearpage

\setcounter{tocdepth}{2}

\tableofcontents

\newpage

\section{Introduction}\label{sec:introduction}

During the course of the swampland approach \cite{Vafa:2005ui} (see \cite{Brennan:2017rbf, Palti:2019pca, vanBeest:2021lhn, Grana:2021zvf, Agmon:2022thq} for reviews) it became clear that not any effective field theory (EFT) 
at low energies can be consistently embedded into quantum gravity in the ultraviolet (UV). 
In fact, the landscape of consistent EFTs appears to be of measure zero. Many swampland arguments can be directly viewed as bottom-up, model-independent constraints on viable EFTs and their UV quantum gravity completions. For instance, there is a prominent and growing body of research exploring the interplay between global symmetries (or lack thereof) \cite{Misner:1957mt, Polchinski:2003bq, Banks:2010zn, Harlow:2018jwu, Harlow:2018tng, Heidenreich:2020pkc, Heidenreich:2021xpr}, topology \cite{McNamara:2019rup, Heidenreich:2020pkc, Heidenreich:2021xpr, McNamara:2021cuo, Blumenhagen:2021nmi, Cvetic:2021sxm, Debray:2021vob, Andriot:2022mri, Blumenhagen:2022bvh, Debray:2023yrs} and black hole physics.

Along these lines, one of the most prominent and most beautiful swampland constraints is the weak gravity conjecture \cite{Arkani-Hamed:2006emk, Harlow:2022ich}, which can be grounded via general black hole
arguments and yet can constrain the mass spectrum of charged particles in quantum gravity.

Another very prominent and important swampland constraint pertains to infinite-distance limits, and is usually dubbed the distance (or duality) conjecture \cite{Ooguri:2006in}. It predicts that moving parametrically large distances along moduli spaces,
the associated EFT must include an infinite tower of asymptotically light states, which become massless in the strict infinite-distance limit. On general grounds, the emergence of an infinite tower rather than a finite number of species can be motivated by the factorization of infinite-distance limits in the language of information geometry \cite{Stout:2021ubb}. The equivalence principle of gravity abhors factorization, and thus gravity must decouple via infinitely many light species in the limit \cite{Stout:2022phm}. In this fashion, the UV cutoff $\Lambda_\text{UV}$ of the corresponding EFT decreases when approaching the infinite distance limits, pushed down by the light tower. This also resonates closely with the magnetic version of the weak gravity conjecture \cite{Arkani-Hamed:2006emk}, whereby limits of small gauge couplings also push the UV cutoff to zero in Planck units. This is not a coincidence, since in string theory gauge couplings typically arise as functions of moduli, and the non-interacting point lies at infinite distance.

From the point of view of string theory, the appearance of infinite towers of light species is a smoking gun for string dualities. When one approaches an infinite-distance limit, the EFT breaks down and is replaced by a dual description, driven by weakly coupled degrees of freedom comprising the light(est) tower. Thus far, the string landscape exhibits a remarkable pattern: the only towers of light species that arise comprise either Kaluza-Klein (KK) modes or the higher-spin excitations of an asymptotically tensionless string. The former reflects the presence of extra dimensions decompactifying, whereas the latter replace the EFT by a full-fledged (but weakly coupled) string theory. Perhaps the most remarkable observation in this context is that this dichotomy persists even when, naively, there are no strings in sight, as in (compactifications of) M-theory or F-theory \cite{Lee:2018urn}. The emergent string conjecture of \cite{Lee:2019wij} encodes this pattern, which in turn only arises because of the precise way string dualities interweave. This scenario, if broadly realized in quantum gravity, completes the picture laid out by the distance conjecture, addressing the nature of the possible towers that become light. Namely, it suggests that only light KK
modes or light string excitations tensionless string arise in general. Many top-down tests of this proposal have been carried out in string theory \cite{Lee:2018urn, Lee:2019wij, Lee:2019xtm}, including non-supersymmetric settings \cite{Basile:2022zee} and excluding potential membrane limits \cite{Alvarez-Garcia:2021pxo}. More general conclusions can be drawn in holographic settings, where fixing internal volumes leaves only higher-spin tensionless limits at infinite distance in the dual conformal manifold \cite{Baume:2020dqd, Perlmutter:2020buo, Baume:2023msm}. Despite these general insights, an all-encompassing bottom-up motivation of this pattern remains elusive. Finding it would be of particular importance, since it would provide strong support to the idea that string theory be the only viable realization of quantum gravity \cite{Montero:2020icj, Bedroya:2021fbu}.

In this paper we aim to provide some bottom-up evidence for the emergent string conjecture resorting to black hole arguments.
Concretely, we will heavily rely on the properties of the species scale \cite{Dvali:2007hz, Dvali:2007wp,Dvali:2009ks,Dvali:2010vm, Dvali:2020wqi}, and closely related black hole arguments to analyze what kind of towers
are allowed in quantum gravity. The species scale $\Lambda_\text{sp}$ can be often thought of as the UV cutoff of an effective theory of gravity, and the EFT description breaks down at this scale. However, according to the original definition it is more properly thought of as the scale at which quantum gravity becomes
strongly coupled. These two scales do not necessarily coincide, since \emph{e.g.} without compactification strings can remain perturbative at the string scale, which is the UV cutoff of the effective field theory.
A crucial property is that the species scale $\Lambda_\text{sp}$ is sensitive to light particles in the EFT, and hence particularly sensitive to light towers of states that emerge in the infinite distance limits. While a finite number of additional light degrees of freedom does not substantially alter the physics within the EFT paradigm, an infinite tower can produce peculiar scalings that relate low-energy scales (the mass scale $m$ of the tower) to the high-energy species scale $\Lambda_\text{sp}$. This is a hallmark of UV/IR mixing, a fundamental feature of quantum gravity.

As we will briefly review, there are at least three possible ways to define the species scale:
first, the perturbative definition relies on the notion of the number $N_\text{sp}$ of particles species below $\Lambda_\text{sp}$ \cite{Dvali:2007hz}, and hence it is well suited in the asymptotic, large distance regime of the moduli space,
where towers of species become light. The resulting scale, which we will denote $\Lambda_\text{sc}$ when distinguishing it from other definitions, encodes the onset of strong gravitational coupling. Secondly, one can closely connect the species scale to the UV cutoff scale $\Lambda_\text{UV}$ appearing in the higher-derivate expansion of the effective gravitational action 
\cite{vandeHeisteeg:2022btw,Cribiori:2022nke,vandeHeisteeg:2023ubh,Cribiori:2023sch,Calderon-Infante:2023uhz,vandeHeisteeg:2023dlw,Castellano:2023aum}. 
In this way, one generically obtains moduli-dependent expressions for the species scale that are globally valid in the entire moduli space, 
in particular also in regimes where the notion of stable particles is lost. The global definition of the species scale also nicely connects it to the theory of modular functions in certain string-theoretic settings. Thirdly, the inverse species scale, \emph{i.e.} the species length $L_\text{sp}$, can be regarded as the minimal size of a black hole that can be consistently described within the EFT description. Therefore, 
these kind of black holes are dubbed minimal black holes.

Actually, the definition of the species scale via minimal black holes has the valuable benefit that the species scale and the associated towers of particles can be connected to fundamental properties of black holes, such as
entropy, mass and temperature. This connection has given rise to the notion of species thermodynamics, which was recently introduced in \cite{Cribiori:2023ffn} and further developed in \cite{Basile:2024dqq, Herraez:2024kux}.
In essence, species thermodynamics associates to a tower of species an entropy ${S}_\text{sp}$ and a temperature $T_\text{sp}$, which is inherited from the entropy and the temperature of the corresponding minimal black hole. A non-trivial link between the two pictures is that the relevant quantities match although computed in different, independent ways.

\begin{figure}[ht!]
    \centering
    \includegraphics[scale=0.5]{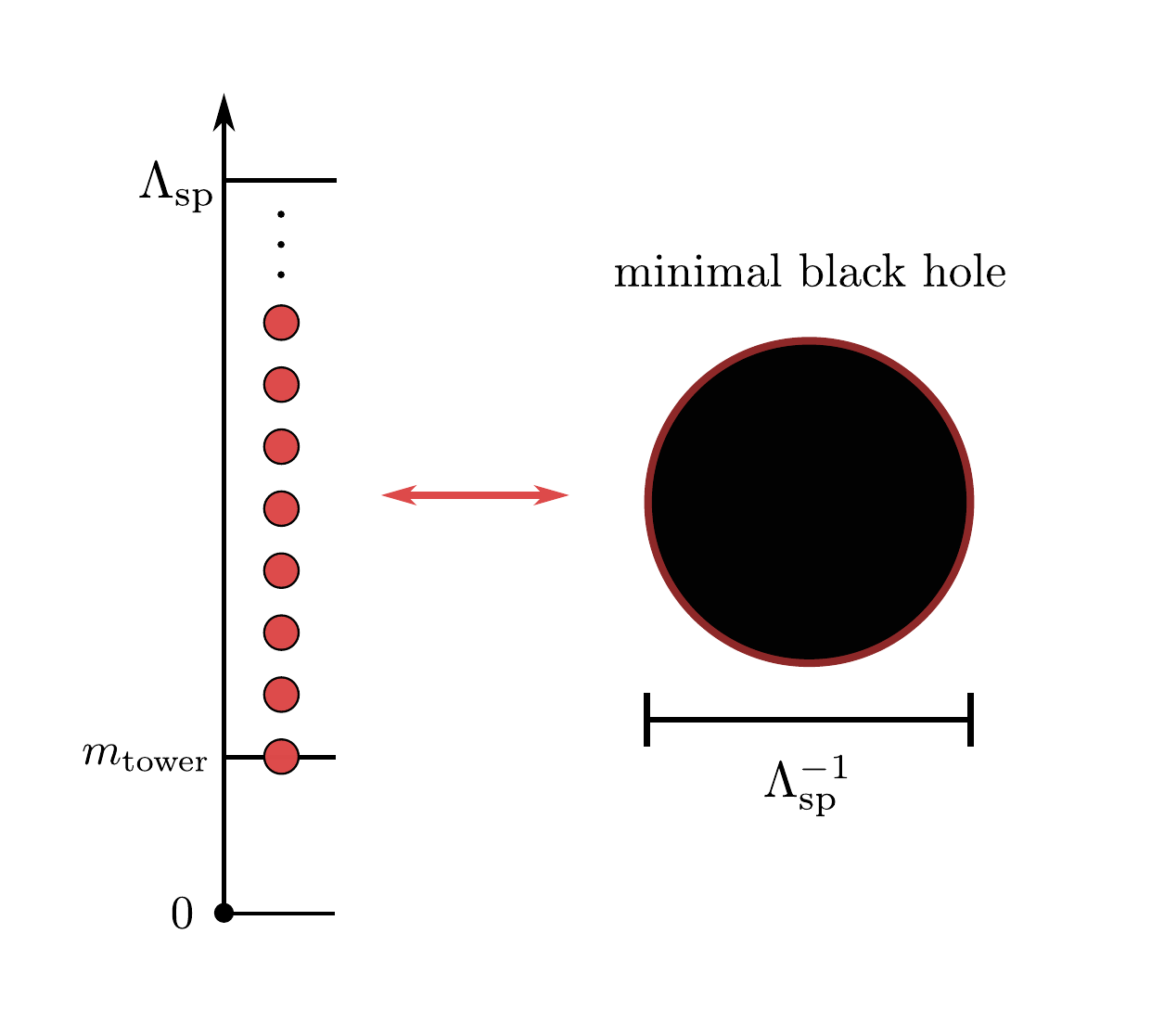}
    \caption{A depiction of the correspondence between (asymptotically) massless towers of species and minimal black holes.}
    \label{fig:species}
\end{figure}

Moreover, in the context of species thermodynamics, minimal black holes can be put in correspondence with (bound states of) a tower of species. This relation is analogous to the $N$-portrait picture \cite{Dvali:2011aa}, where general Schwarzschild-like black holes can be effectively viewed as bound states of species and gravitons. In this way one is also able to derive the species energy $E_\text{sp}$ as the total mass of the species bound states. 
The fundamental connection that we shall exploit is depicted in fig. \ref{fig:species}.

Within an EFT, minimal black holes obey the well-known general thermodynamic laws between black entropy, energy (i.e. mass) and temperature. As we will discuss and demonstrate in our paper, imposing these relations 
on the thermodynamic quantities ${S}_\text{sp}$, ${T}_\text{sp}$ and to ${E}_\text{sp}$, leads to non-trivial constraints on the possible spectra of the associated towers of species as constituents of minimal black hole bound states.
So as we shall see in detail, one of the main observations in this paper is that not all towers of species are consistent with the laws of
species thermodynamics and yield the correct expression for $E_\text{sp}$. 
We will show that minimal black holes corresponding to light KK or light string towers indeed satisfy these constraints, thus providing consistent
solutions of species thermodynamics in this sense. 
 
The converse, bottom-up question is equally or perhaps even more relevant. Namely, we ask whether KK and string towers are the only possible solutions of the above thermodynamic constrains, or if there can exist also other light towers of states which build minimal black holes.
By studying the general restrictions and also some particular tower spectra, which violate these constraints, we will shed some light on which kind of species spectra are consistent from the thermodynamic point of view. Our basic conclusion will be that the bottom-up black hole arguments point towards the emergent string conjecture, sorting the possible towers into two different sectors: KK-like towers and string-like towers. 
In other words, from a purely bottom-up approach, the only tower which can exhibit a transition with a black hole are the Kaluza-Klein towers and string towers. Hence, the Emergent String Conjecture can be expressed in terms of a transition between black holes and towers of species. 

The paper is organized as follows: in section \ref{sec:relevant_scales} we will review different ways in which the relevant UV scales in quantum gravity can be defined. As we will point out, there are some subtle differences between the perturbative definition of the species scale and the UV cutoff suppressing the higher-derivative expansion. We address how these differences can be interpreted and resolved, as this is also relevant for our discussion about the emergent string conjecture. In section \ref{sec:top-down}, following \cite{Cribiori:2023ffn}, we show that the KK towers and the string towers indeed satisfy the required thermodynamic relations, which means that they can be used to build minimal black hole bound states in a consistent way. In section \ref{sec:bottom-up} we turn the argument around, namely by asking from the bottom-up perspective which kind of general tower spectra do satisfy 
the thermodynamic relations. As we shall see, these requirements impose non-trivial constraints on the form of the species mass levels as well as on their possible degeneracies. Specifically, we shall derive a consistency relation between mass spectra and degeneracies of towers of species. Hence, requiring the possibility of a transition between black holes and towers of species restricts the classes of towers dictated by the emergent string conjecture. All in all, the overall lesson is that a consistent match with the thermodynamics of minimal black holes rules out certain classes of towers, essentially leaving KK and string towers as the only option for a tower of species. Subsequently, we shall present a case-by-case analysis of various types of towers and the source of their (in)consistency.

\section{Relevant scales in quantum gravity}\label{sec:relevant_scales}

There are a number of scales that can be defined in EFTs of quantum gravity. The species scale, as traditionally defined \cite{Dvali:2007hz, Dvali:2007wp}, is the scale at which quantum gravity effects cease to be negligible. Namely, it is the scale at which the effective quantum coupling between gravitons becomes $\mathcal{O}(1)$. When ambiguities may arise, we dub this the ``strong coupling'' scale $\Lambda_\text{sc}$, in order to distinguish it from other relevant scales. The strong coupling scale can be obtained along the lines of \cite{Dvali:2007wp} (see also \cite{Castellano:2022bvr, Blumenhagen:2023yws}) considering the effective coupling in the 1-loop graviton propagator. For weakly interacting species, the leading contribution is proportional to the number $N_\text{sp}(\Lambda_\text{sc})$ of light species. Imposing that the effective coupling be of order one yields the (implicit) parametric relation
\begin{equation}\label{eq:species_scale_definition}
    \Lambda_\text{sc} = \frac{M_\text{pl}}{N_\text{sp}^{\frac{1}{d-2}}} \, ,
\end{equation}
which may be solved for $\Lambda_\text{sc}$. In the following, as well as in the remainder of the paper, we shall work in $d$-dimensional Planck units $M_\text{pl} = 1$ unless otherwise specified. We also mostly work with leading-order asymptotics in infinite-distance limits $m_\text{tower} \ll M_\text{pl}$, neglecting order-one multiplicative factors when they are irrelevant.

Another relevant scale is the UV cutoff of the effective field theory, $\Lambda_\text{UV}$. This is the \emph{typical} scale suppressing higher-derivative corrections in a (Wilsonian) effective action whose gravitational sector has an asymptotic expansion of the schematic form
\begin{equation}
    S_\text{eff} \sim \frac{M_\text{pl}^{d-2}}{2} \int d^d x \, \sqrt{-g} \left( R + \sum_n \frac{c_n}{\Lambda_\text{UV}^{2n-2}} \, O_n(g,\text{Riem},\nabla) \right) \, .
\end{equation}
By \emph{typical} we mean that $\Lambda_\text{UV}$ is the parametric scale suppressing almost all such corrections, namely it is defined in such a way that almost all Wilson coefficients $c_n$ be of order one in the EFT limit. This is guaranteed if the theory contains no dimensionless parameters or moduli, but in their presence it is \emph{a priori} non-trivial. Indeed, it has been argued that at most finitely many Wilson coefficients may be fine-tuned \cite{Heckman:2019bzm}, and their moduli dependence can be constrained by rather general EFT considerations \cite{vandeHeisteeg:2022btw}.

One is naturally led to ask whether these two scales coincide. In order to investigate this question, we consider simple settings in which one can estimate both scales $\Lambda_\text{sc} \, , \, \Lambda_\text{UV}$ in terms of a single parameter. For generic values of this parameter, one expects all scales to be of the order of $M_\text{pl}$, since there is no other option. This is indeed what is found at \emph{desert points} in moduli spaces \cite{Long:2021jlv, vandeHeisteeg:2022btw}. However, for parametrically small values one expects infinite light towers of species to arise \cite{Ooguri:2006in}. This phenomenon is supported by a compelling bottom-up motivation, based on information theory \cite{Stout:2021ubb}. Namely, small limits of physical control parameters\footnote{In quantum gravity, any such parameter is widely expected to be the vacuum value of some dynamical field. This is true, for instance, in string perturbation theory and AdS/CFT holography. It is also connected to the absence of $(-1)$-form symmetries \cite{McNamara:2020uza}.} encode infinite-distance limits in the information metric. These limits invariably lead to \emph{factorization} of expectation values, which in turn seems to require infinite towers of species when gravity is involved \cite{Stout:2022phm}. This enticing observation can be ultimately traced back to the equivalence principle, whereby gravity couples to everything in the same fashion, and its coupling cannot be tuned since it depends on the energy-momentum distribution of matter.

In light of the observed pattern that infinite-distance limits in the string landscape only seems to yield light KK towers or excitations of asymptotically tensionless strings \cite{Lee:2019wij}, we consider weakly coupled strings in flat $d$-dimensional spacetime and KK towers driven by a single overall internal $p$-dimensional volume, $\text{Vol} \propto m_\text{KK}^{-p}$. In the former case, the relevant limit is $g_s \ll 1$, whereas in the latter case it is $m_\text{KK} \ll 1$. Notice that the former limit can be recast as $M_s \ll 1$, on account of the relation
\begin{equation}
    M_\text{pl} = M_s \, g_s^{-\frac{2}{d-2}}
\end{equation}
between the string coupling $g_s$, the Planck mass and the string scale $M_s$. In other words, in Planck units weakly coupled strings are light, since their excitation modes arrange in a tower whose typical mass scale is $M_s$. However they are not light relative to the UV cutoff of the EFT, a fact which will play an important role in our analysis.

The procedure described above to compute $\Lambda_\text{sc}$ yields, parametrically,
\eq{
    & \Lambda_\text{sc}^\text{string} = M_s \log \frac{M_\text{pl}}{M_s} = M_s \log \biggl(\frac{1}{g_s}\biggr)^{\frac{2}{d-2}}\, , \\
    & \Lambda_\text{sc}^\text{KK} = m_\text{KK}^{\frac{p}{p+d-2}} = M_{\text{pl},d+p} \, .
}
Here, $M_{\text{pl},d+p}$ is the Planck scale of the higher-dimensional theory. This result for weakly coupled strings is somewhat puzzling, as commented in \cite{Blumenhagen:2023yws}. Here we will frame this subtlety from a different perspective, but first let us discuss the higher-derivative scale $\Lambda_\text{UV}$ which controls the EFT expansion.

\subsection{Typical higher-derivative scales}\label{sec:typical_HD_scales}

To begin with, the higher-derivative scale suppressing EFT corrections for weakly coupled strings is simply the string scale itself, $\Lambda_\text{UV} = M_s$. This is apparent from the mechanism with which string perturbation theory matches low-energy observables, either via scattering amplitudes or (at tree level) Weyl anomaly cancellation on the worldsheet. String loop corrections are suppressed in the limit we are considering. It is worth noting that this UV cutoff can be motivated by bottom-up arguments as well \cite{Afkhami-Jeddi:2018apj}.

For the simple decompactification limit we described, the situation is \emph{a priori} not so clear, since there is no obvious UV completion to refer to. Still, one can identify two sources of higher-derivative corrections to the $d$-dimensional effective action. One stems from dimensionally reducing higher-derivative terms in the $(d+p)$-dimensional effective action, while the other arises integrating out KK fluctuations\footnote{The interplay between these two types of corrections has been investigated in \cite{Arfaei:2023hml}.}. Absent any further parametric limits, the former corrections are suppressed by the $(d+p)$-dimensional Planck scale $M_{\text{pl},d+p}$ in the higher-dimensional theory. This is also true in string theory, since we are not taking $g_s \to 0$ in the higher-dimensional theory and thus $M_s \sim M_{\text{pl},d+p}$ parametrically. Dimensional reduction then yields terms of the type
\begin{align}
    \int d^{d+p}x \, \sqrt{-g_{d+p}} \, \frac{O_n}{M_{\text{pl},d+p}^{2n-d-p}} & \sim \int d^d x \, \sqrt{-g} \, \frac{O_n}{M_{\text{pl},d+p}^{2n-d-p}m^p} = M_\text{pl}^{d-2} \int d^d x \, \sqrt{-g} \, \frac{O_n}{\Lambda_\text{sc}^{2n-2}}
\end{align}
identifying $\Lambda_\text{UV}^{\text{tree}} = \Lambda_\text{sc}$ for these tree-level contributions. In order to estimate the quantum contribution arising from KK loops, one can observe that the dominant diagram generating $O_n$ is expected to contain one loop of KK modes. Applying dimensional analysis requires reinstating the Planck scale $M_\text{pl}^{d-2}$. The contribution arising from a single KK loop leads to a sum over the tower, and thus to an effective action of the form
\begin{equation}
    \int d^d x \, \sqrt{-g} \left( M_\text{pl}^{d-2} \, R + \sum_n c_n \, \sum_{\mathbf{k} : \abs{\mathbf{k}} \leq N_\text{sp}^{1/p}} (m_\text{KK}\abs{\mathbf{k}})^{d-2n} \, O_n \right) \, .
\end{equation}
When the sum diverges in the infinite-distance limit, the coefficient of $O_n$ can then be estimated as\footnote{In principle, if the tower contains higher-spin fields the interactions may introduce a different asymptotic scaling of the couplings (see \emph{e.g.} \cite{Afkhami-Jeddi:2018apj}). Higher-spin towers cannot be interpreted in terms of a higher-dimensional QFT and are subjected to very strong unitarity constraints \cite{Afkhami-Jeddi:2018apj, Arkani-Hamed:2020blm, Geiser:2022exp, Cheung:2022mkw, Cheung:2023adk, Cheung:2023uwn}, which may uniquely select the string spectrum. We shall discuss this point in more detail in section \ref{sec:complementary_arguments}.}
\begin{equation}\label{eq:KK_loop_estimate}
    (\Lambda_\text{UV}^{\text{loop}})^{2-2n} \equiv \sum_{\mathbf{k} : \abs{\mathbf{k}} \leq N_\text{sp}^{1/p}} (m_\text{KK}\abs{\mathbf{k}})^{d-2n} \sim m_\text{KK}^{d-2n} \, N_\text{sp}^{1+\frac{d-2n}{p}} \sim \Lambda_\text{sc}^{2-2n} \, .
\end{equation}
Otherwise, when the sum converges, the one-loop EFT estimate cannot be applied. This simple argument\footnote{More sophisticated versions of this argument are presented in \cite{Blumenhagen:2023yws,Calderon-Infante:2023uhz}.} indicates that, indeed, for decompactification limits the UV cutoff is the strong coupling scale, $\Lambda_\text{UV} = \Lambda_\text{UV}^\text{tree} = \Lambda_\text{UV}^\text{loop} = \Lambda_\text{sc}$. As we have remarked, this is the typical scale suppressing higher-derivative operators, but in principle there can be finitely many exceptions; see \emph{e.g.} \cite{Green:2010wi} for examples in the strong-coupling limit of type IIA superstrings in ten dimensions. It would be interesting to extend these considerations to more general KK limits where multiple (pseudo-)moduli play a role and significant anisotropies may arise\footnote{See, however, \cite{Collins:2022nux} for constraints on anisotropies grounded in AdS/CFT holography.}. For a general tower of states $\{m_\mathbf{s}\}$ one can adapt the above estimate, which results in a family $\Lambda_\text{UV}^{(n)}$ of \emph{upper bounds} to the true cutoff. In general,
\begin{equation}
    \Lambda_\text{UV} \leq \Lambda_\text{sc} \, , \, \inf_n \Lambda_\text{UV}^{(n)} \, .
\end{equation}
If for some tower the estimate leads to $\inf_n \Lambda_\text{UV}^{(n)} \ll \Lambda_\text{sc}$, one can conclude that the tower is not light with respect to the UV cutoff. Otherwise, no clear diagnosis can be drawn asbent more data about a UV completion.

All in all, at least for these simple settings, only for weakly coupled strings one finds $\Lambda_\text{sc} \gg \Lambda_\text{UV}$ by a logarithmic factor. We do not find this surprising: the effective coupling of strings remains small at energies of the order $M_s$. To wit, recall that $2\to 2$ closed-string scattering at high center-of-mass energies \cite{Gross:1987kza, Gross:1987ar} $\sqrt{s} \gg M_s$ and fixed angle is encoded in tree-level amplitudes of the form $g_s^2 \, \exp(-s/M_s^2)$. The loop corrections scale in a similar fashion \cite{Gross:1987kza} as $g_s^{2+2g} \, \exp(-\frac{s}{(g+1)M_s^2})$, which suggests that the onset of the non-perturbative regime is reached at scales $\sqrt{s} \sim M_s \, \sqrt{\log g_s^{-2}}$. This scale is parametrically larger than $M_s$, where strings remain weakly coupled, and it is parametrically smaller than $\Lambda_\text{sc}$, consistently with the fact that the graviton computation \emph{a priori} provides an upper bound for the strong coupling scale. A more sophisticated treatment requires estimating the full perturbative series via \emph{e.g.} Borel resummation \cite{Mende:1989wt}\footnote{See also 
\cite{Dvali:2014ila} for a study of black hole formation in $2\to N$ scattering, and \cite{Bedroya:2022twb} for a recent review of the black hole/string transition in this context.}. This gives amplitudes scaling as $\exp(-\sqrt{\frac{s}{M_s^2 \, \log \, g_s^{-2}}})$ in the relevant regime with $g_s \to 0$. The same strong coupling scale appears as a characteristic scale in this expression.

\section{Top-down: minimal black holes from light towers}\label{sec:top-down}

In this section we review the computation of species thermodynamic quantities for the two well-known cases of weakly coupled string excitations and simple KK towers, controlled by a single mass scale $m_\text{KK}$. From here onward, we shall parametrize towers of species in terms of an effective level $n$, such that the mass
\begin{equation}\label{eq:general_mass}
    m_n = m \, f(n)
\end{equation}
is controlled by a single typical scale $m$. Each level has a degeneracy $d_n$. Since we are interested in leading-order (and occasionally subleading) effects in thermodynamics, \emph{a priori} multiple equivalent parametrizations could exist. For KK towers, this is shown \emph{e.g.} in \cite{Castellano:2022bvr}. Let us emphasize that towers of this type are, in a sense, ``gapless'' -- there is only one characteristic mass scale controlling the spectrum, and the whole tower becomes massless in the limit $m \to 0$. A parametric gap inside the tower would modify our considerations, but if it were separating only a finite number of states from the rest of the tower one could include these states in the EFT from the outset. In the following, as well as in the remainder of the paper, we shall work with this types of towers.

\subsection{Black hole picture for species towers}\label{sec:BH_picture}

Before introducing the thermodynamics of species, let us recall their connection with minimal black holes. In the spirit of \cite{Cribiori:2023ffn}, the smallest black hole that can be reliably described within a gravitational EFT has a size $R_\text{S} = \Lambda_\text{sp}^{-1}$. Focusing on Schwarzschild-like minimal black holes, their mass, entropy and temperature are thus given by (up to numerical factors)
\begin{align}\label{eq:BH_mass}
    M_\text{BH} \sim \Lambda_\text{sp}^{3-d} \,,\quad
    S_\text{BH} \sim \Lambda_\text{sp}^{2-d} \, , \quad
    T_\text{BH} \sim \Lambda_\text{sp}
\end{align}
up to numerical factors.
As introduced above, and explained in more detail in the following, we will consider towers of states whose masses are determined by a level $n$ and mass scale $m$. The spectrum of masses is of the general form 
as given in eq. \eqref{eq:general_mass} for an unbounded increasing function $f$, and each level has a degeneracy $d_n$. In this setting, minimal black holes are built as bound states of precisely one of each species up to the maximum level $N$, defined by
\begin{equation}\label{eq:max-level_species_scale}
    m_N = \Lambda_\text{sp} \, .
\end{equation}
This prescription aligns with the $N$-portrait picture of \cite{Dvali:2011aa}, and yields the correct thermodynamic scalings for minimal black holes in top-down examples \cite{Cribiori:2023ffn}. At any rate, $\mathcal{O}(1)$ deviations from this prescription would yield $\mathcal{O}(1)$ factors in thermodynamic quantities.

As a result, the leading-order scaling of the mass of the resulting black hole ought to be
\begin{align}
    M_\text{BH} = E_\text{sp} = \sum_{n \leq N} d_n \, m_n \, .
\end{align}
We remark that we do not include any binding energy nor backreaction, since in this
black hole/species correspondence the relevant tower of light species is weakly coupled
approaching infinite distance in moduli space. These effects could be included computing
thermodynamic quantities along the lines of [70]. Whether this construction is compatible with the scaling in eq. \eqref{eq:BH_mass} turns out to be non-trivial.  We remark that we do not include any binding energy nor backreaction, since in this black hole/species correspondence the relevant tower of light species is weakly coupled approaching infinite distance in moduli space. These effects could be included computing thermodynamic quantities along the lines of \cite{Kolekar:2010py}. Whether this construction is compatible with the scaling in eq. \eqref{eq:BH_mass} turns out to be non-trivial. There are in fact examples of towers that we will exclude on these grounds. As a general remark, our thermodynamic expressions are only valid for integer degeneracies. The leading-order results are unaffected, but in principle the subleading terms could change. We leave a more detailed analysis of this subtlety to future work, but we note that our results fit into a sensible pattern which only depends on the qualitative structure of these corrections.

\subsection{Microcanonical species entropy}\label{sec:microcanonical_species_entropy}

We now present in more detail the construction of the species ensemble. Fixing the species energy $E_\text{sp} \sim \Lambda_\text{sp}^{3-d}$ to the mass of a minimal black hole defines a microcanonical ensemble, whose microstates comprise tuples $\{k_\mathbf{s}\}$ of occupation numbers of each state and species. The species entropy $S_\text{sp} = \log D(E_\text{sp})$ is then the microcanonical entropy of this ensemble. In order to compute it, we consider the auxiliary partition function
\begin{equation}
    Z(q) = \sum_M D(M) \, q^M = \prod_{n\leq N} \frac{1}{(1 - q^{f(n)})^{d_n}} \, ,
\end{equation}
which counts how many microstates have total energy $m \, M$. For the minimal black hole one has
\begin{equation}\label{eq:minimal_BH_M_value}
    M = \frac{E_\text{sp}}{m} = \frac{E_\text{sp}}{\Lambda_\text{sp}^{3-d}} \, N_\text{sp} \, f(N) \sim N_\text{sp} \, f(N) \gg N_\text{sp}
\end{equation}
whenever a parametrically large number of levels are below the species scale. The entropy can then be read off from the leading asymptotics of\footnote{Similarly to the case of the energy, strictly speaking one should work with integer $f$ to count powers of $q$, but the leading-order result is unaffected by this subtlety.}
\begin{equation}
    D(M) = \frac{1}{2\pi i} \oint \frac{dq}{q^{M+1}} \, Z(q) \, .
\end{equation}
For $M\gg 1$, the saddle point equations for the exponent $-(M+1)\log q - \log Z(q)$ single out the saddle $q_* = 1 - \epsilon$ with $\epsilon \ll 1$. Expanding, one obtains
\begin{equation}
    q_* \overset{M \gg N_\text{sp}}{\sim} 1 - \, \frac{N_\text{sp}}{M}
\end{equation}
at leading order. As a result, including the first subleading correction arising from the Gaussian integral, when the dust settles one arrives at
\begin{equation}\label{eq:anyenergy_species_entropy}
    S(M) \overset{M \gg N_\text{sp}}{\sim} N_\text{sp} + N_\text{sp} \, \log \frac{M}{N_\text{sp}} - \sum_{n \leq N} d_n \, \log f(n) - \, \frac{1}{2} \, \log \frac{M^2}{N_\text{sp}} \, .
\end{equation}
From eq. \eqref{eq:anyenergy_species_entropy}, varying with respect to the energy one finds the species temperature
\begin{equation}
    T_\text{sp} = m \left(\frac{\partial S}{\partial M}\right)^{-1}_{M = N_\text{sp} f(N)} \sim m \, f(N) = \Lambda_\text{sp} \, .
\end{equation}
When specializing to minimal black holes, one expects the species entropy, namely the entropy for $M \sim N_\text{sp} \, f(N)$ according to eq. \eqref{eq:minimal_BH_M_value}, to be given by
\begin{equation}\label{eq:pre-general_species_entropy}
    S_\text{sp} \sim N_\text{sp} + \sum_{n \leq N} d_n \, \log \frac{f(N)}{f(n)} - \, \frac{1}{2} \log(N_\text{sp} \, f(N)^2) \, .
\end{equation}
However, as we shall see, in some spurious cases the subleading corrections to the asymptotics $M \sim N_\text{sp} \, f(N)$ can be of the same order of other terms in the above formula. In order to account for these, we write $E_\text{sp} \equiv \gamma \Lambda_\text{sp}^{3-d} \left( 1 + \epsilon \right)$, with $\epsilon \ll 1$, and define\footnote{The prefactor $\gamma$ only changes the constant in front of $N_\text{sp}$ in the final result. We define $M$ in this fashion for simplicity, but also to match the combinatorial counting of \cite{Blumenhagen:2023yws}.} $M = (1 + \epsilon) \, N_\text{sp} \, f(N)$. As a result, the full asymptotics of the species entropy becomes
\begin{equation}\label{eq:general_species_entropy}
    S_\text{sp} = N_\text{sp} (1 + \log(1+\epsilon)) + \sum_{n \leq N} d_n \, \log \frac{f(N)}{f(n)} - \, \frac{1}{2} \log(N_\text{sp} \, f(N)^2) + \mathcal{O}(1) \, .
\end{equation}
In the remainder of this section we will use these expressions to show that simple KK towers and string towers satisfy $S_\text{sp} \sim N_\text{sp}$ as the leading-order contribution, as expected from the species thermodynamics picture of \cite{Cribiori:2023ffn}.

\subsection{From Kaluza-Klein towers to minimal black holes}\label{sec:KK_to_BH}

We begin by considering simple KK towers parametrized by a single mass scale $m_\text{KK} \ll 1$. KK modes for an internal $p$-torus are characterized by integer quantum numbers $\mathbf{k} \in \mathbb{Z}^p$, and masses $m_{\mathbf{k}} = m_\text{KK} \, \abs{\mathbf{k}}$. For our purposes, such towers can be effectively described in terms of a level $n$ with constant degeneracy and a power-like mass spectrum \cite{Castellano:2022bvr, Castellano:2023qhp} 
of the form\footnote{An equivalent parametrization of the tower is given by masses $f(n)=n$ and degeneracies $d_n=n^{p-1}$.}
\begin{equation}
f(n) = n^{\frac{1}{p}} \, , \qquad m = m_\text{KK}
\end{equation}
in the notation of eq. \eqref{eq:general_mass}. The exponent encodes the number $p$ of compactified extra dimensions. The highest level of the tower allowed by the weakly coupled description, denoted as $N$, is defined by eq. \eqref{eq:max-level_species_scale}. Since the degeneracy $d_n = \beta$ remains constant, for instance a spin degeneracy, one finds
\begin{equation}
\Lambda_\text{sp} = m_N = N^{\frac{1}{p}} m_\text{KK} \, .
\end{equation}
This simplifies the calculation of the total number of species within the tower, denoted as $N_\text{sp}$, which is just given by
\begin{equation}
N_\text{sp} = \beta N \, .
\end{equation}
To compute the species energy, which ought to match the mass of a minimal black hole comprising exactly one of each species, we sum the masses associated with each level in the tower. One finds
\begin{equation}
E_\text{sp} = \sum_{n=0}^N n^{\frac{1}{p}} \, m_\text{KK} = m_\text{KK} \, H_N^{(-\frac{1}{p})} \equiv m_\text{KK} \left[-\zeta\left( -\frac{1}{p}, N+1\right) + \zeta\left(-\frac{1}{p}\right) \right],
\end{equation}
where $\zeta(s, q)$ is the Hurwitz zeta function and $H_n^{a}$ denotes the generalized harmonic numbers. Our focus lies in the behavior of these quantities as $N$ grows large. In the large-$N$ limit, the asymptotics of the Hurwitz zeta function enable us to simplify the above expression according to
\begin{equation}
E_\text{sp} = m_\text{KK} \, N^{1/p} \left(\frac{p}{p+1} N +\frac{1}{2} + \mathcal{O}\left(\frac{1}{N} \right) \right).
\end{equation}
This can be further recast as
\begin{equation}\label{GeneralEnergyRel}
E_\text{sp} \sim \frac{p}{p+1} \, \Lambda_\text{sp}^{3-d} + \frac{1}{2}\, \Lambda_\text{sp} \, .
\end{equation}
We see that this relation bears resemblance to the correlation between the mass of a black hole $M_\text{BH}$ and its Schwarzschild radius $R_\text{S}$, which is $M_\text{BH} \sim R_\text{S}^{d-3}$ for large radii. Since $N_\text{sp} = N$ in this case, using eq. \eqref{eq:general_species_entropy}, one obtains for entropy
\begin{equation}
    S_\text{sp} \sim  N + \frac{1}{p} \, N \, \log N -  \frac{1}{p}\sum_n \log n \sim \frac{p+1}{p} \, N \propto \Lambda_\text{sp}^{2-d} \, .
\end{equation}
One can also compute the subleading correction $- \frac{p+3}{2p} \log N$ from eq. \eqref{eq:anyenergy_species_entropy}, with the next correction being $\mathcal{O}(1)$. Finally, we turn to the temperature of the tower, finding
\begin{equation}
T_\text{sp} = \dfrac{dE_\text{sp}}{dS_\text{sp}} = \frac{d-3}{d-2} \, \Lambda_\text{sp} + \mathcal{O}(\Lambda_\text{sp}^{d-1}) \, .
\end{equation}
In conclusion, we find the KK towers are fully consistent with the thermodynamics laws in eq. \eqref{eq:BH_mass} required for minimal black holes. In other words, minimal black holes can be viewed as bound states KK towers in a consistent way.

\subsection{From string towers to minimal black holes}\label{sec:string_to_BH}

We now discuss weakly coupled string towers. In this case, the mass spectrum is typically written as
\begin{eqnarray}
m_n = \sqrt{n-1} \, M_s
\end{eqnarray}
in terms of the string scale $M_s$. For bosonic strings in flat spacetime $n\geq 0$ includes the tachyon. Moreover, the degeneracy of states $d_n$ is asymptotically exponential at large excitation levels $n$, and can be expressed as $d_n = \frac{\beta}{n^\kappa}e^{\sqrt{n}}$\footnote{A similar expression for a gas of $p$-branes compactified on $(S^1)^p \cross \mathbb{R}^{D-p}$ can be obtained via semiclassical methods \cite{Alvarez:1991qs, KalyanaRama:2003cz}, resulting in the exponential behavior $\exp N^{\frac{p}{p+1}}$. As noticed in \cite{Alvarez:1991qs}, this reproduces the exact scaling of black hole entropy only for $p \to \infty$.}.

To quantify the number of species $N_\text{sp}$ under the strong coupling scale, we need to account for this exponential effect. We approximate $N_\text{sp}$ in the limit of large $N_\text{sp}$ as
\begin{equation}
N_\text{sp} = \sum_{k=0}^{N} d_k = \beta \sum_{k=0}^{N} e^{\sqrt{k}}k^{-\kappa} \sim 2 \beta \left(\sqrt{N}+\left(\kappa -\frac{3}{4}\right)\right) e^{\sqrt{N}} N^{-\kappa} \, ,
\end{equation}
where $N$ is the maximum excitation level such that the mass remains below $\Lambda_\text{sp}$. Namely,
\begin{equation}
\Lambda_\text{sp} = m_{N} \overset{N \gg 1}{\sim} \sqrt{N} M_s \, .
\end{equation}
This definition enables us to determine $N$ in terms of the string mass $M_s$ using the $W_{-1, 0}$ Lambert function for every $d$ and $\kappa$. Asymptotically,
\begin{equation}
\sqrt{N} = (d-2) \log{\frac{1}{M_s}} + \mathcal{O}\left(\log{\log{\left(\frac{1}{M_s}\right)}}\right) \, .
\end{equation}
Consequently, we can express the species scale in terms of $M_s$ as
\begin{equation}\label{stringMultCorrection}
\Lambda_\text{sp} = (d-2) \, M_s \log{\frac{1}{M_s}} + \mathcal{O}\left(\log\left(\frac{1}{M_s}\right)\right).
\end{equation}
In this case, as we have remarked in section \ref{sec:relevant_scales}, $\Lambda_\text{sp} = \Lambda_\text{sc}$ is a(n upper bound to the) strong coupling scale for gravitational effects. The UV cutoff of the EFT is instead $\Lambda_\text{UV} = M_s$.

We proceed to calculate the species energy, which involves the weighted sum of string masses
\begin{equation}\label{eq:string_species_energy}
E_\text{sp} = \Lambda_\text{sp}^{3-d} \left( 1 + \frac{1}{d-2} \frac{1}{\log{(\Lambda_\text{sp})}} + \mathcal{O}\left( \frac{1}{\log^2{\Lambda_\text{sp}}}\right) \right) \equiv \gamma(\log{\Lambda_\text{sp}}) \, \Lambda_\text{sp}^{3-d} \, .
\end{equation}
All in all, also in this scenario we uncover the same relationship as eq. \eqref{eq:BH_mass} between the ensemble energy and the minimal black hole mass describable within the EFT, except for a subtlety. The main difference in eq. \eqref{eq:string_species_energy} with respect to the expected result is a multiplicative logaritmic correction arises from eq. \eqref{stringMultCorrection}, so that the effective prefactor $\gamma$ is not constant upon including subleading terms, but rather depends logarithmically on the species scale. This peculiarity pertains not only to the energy, as we shall now see, but we may ascribe it to the fact that the tower of string excitations is not light with respect to the UV cutoff. Moreover, taking into account also the second correction to the species energy, one finds that eq. \eqref{eq:general_species_entropy} gives\footnote{This correction may be modified taking into account the exact integer degeneracies of the string spectrum.}
\begin{equation}\label{eq:string_entropy}
    S_\text{sp} \sim N_{\text{sp}} \left(1 + \mathcal{O}{\left(\frac{\log \log N_\text{sp}}{\log^2 N_\text{sp}}\right)} \right) \, ,
\end{equation}
which is consistent with the expected behavior of black hole entropy at leading order. Finally, we determine the temperature of this tower,
\begin{eqnarray}\label{eq:string_temperature}
T_\text{sp} = \dfrac{dE_\text{sp}}{dS_\text{sp}} \sim \frac{d-3}{d-2} \, \Lambda_\text{sp} + \mathcal{O}\left(\Lambda_\text{sp}^{d-1}\right)
\end{eqnarray}
Of notable interest is the relation between this temperature and the species scale in terms of the string length, which reads
\begin{eqnarray}
T_\text{sp} \sim M_s \log{M_s^{-1}} \, .
\end{eqnarray}
This temperature corresponds to the Hagedorn temperature up to a multiplicative logarithm correction. It encodes the point at which the black hole analogy loses its semiclassical description, leading to a transition between highly excited stringy degrees of freedom and black hole microstates.

The multiplicative logarithmic corrections to the species energy and entropy may appear puzzling at first glance, but they can be ascribed to the fact that the string tower is not light with respect to the UV cutoff $\Lambda_\text{UV} = M_s$, and thus the EFT expansion of the mass-radius relation, which would not contain these contributions, cannot be trusted. Indeed, the transition between black holes and string states is expected to take place at the scale \cite{Bedroya:2022twb} 
\begin{equation}\label{eq:BH_transition}
    \frac{M_s}{g_s^2} = M_s \left( \frac{M_\text{pl}}{M_s} \right)^{d-2} = M_s^{3-d} \gg \Lambda_\text{UV} \, ,
\end{equation}
well outside of the regime of the effective field theory. This seems deeply tied to the UV/IR mixing manifested by classicalization, \emph{i.e.} the formation of large black holes from high-energy scattering. Once the black hole forms, eq. \eqref{eq:BH_transition} is to be reinterpreted as an infrared scale $R_\text{S} \gg \ell_\text{pl}$. Interestingly, this regime appears reachable within (resummed estimates of) string perturbation theory \cite{Bedroya:2022twb}. This indicates that the logarithms in eq. \eqref{stringMultCorrection}, and thus in eqs. \eqref{eq:string_species_energy} and \eqref{eq:string_entropy}, are an artifact of the EFT analysis. Actually, this is corroborated by the considerations in \cite{Dvali:2009ks}: not all states in the tower participate equally in a scattering or black hole formation process. Indeed, in \cite{Dvali:2009ks} it is emphasized that the effective couplings $\gamma_n$ of higher-spin strings excitations to external states of a scattering process are suppressed at large exchanged spin. To see this, these couplings can be read off of the pole expansion of scattering amplitudes. As a simple example, the Veneziano amplitude decomposes according to
\begin{equation}
    A(s,t) \sim \frac{\gamma_n}{s - m_n^2} \, ,
\end{equation}
with
\begin{equation}
    \gamma_n \sim \frac{(\alpha' u)^n}{n!} \, ,
\end{equation}
where $n+1$ is the highest spin exchanged at the resonance. Because the effective coupling of the intermediate higher-spin excitations is suppressed for large $n$, although the degeneracy of states grows exponentially with their mass not all states can be effectively excited during a scattering or black-hole formation process. As a result, the effective number of species that contributes is $N_\text{eff} = g_s^{-2}$. This produces an effective species scale $M_\text{pl} / N_\text{eff}^{\frac{1}{d-2}} = M_s$. This calculation, in contrast to eq. \eqref{stringMultCorrection}, produces the correct UV cutoff for the EFT arising from weakly coupled strings without the logarithmic contribution. After taking into account the effective counting of the species, then the species entropy and energy have no spurious terms.
Analogous considerations can be made for KK towers, as we explain in more detail in appendix \ref{app:KK_BH}.

\section{Bottom-up: light towers from minimal black holes}\label{sec:bottom-up}

This section delves into the possibility of constructing minimal black holes via towers of species. We aim to shed some light on a fundamental question, namely whether it is always feasible to build a tower such that its energy profile conforms,
according to eq. \eqref{eq:BH_mass}, to the following expression, 
\begin{equation}\label{eq:species_power_correction}
E_\text{sp} = \gamma \, \Lambda_\text{sp}^{3-d} + \mathcal{O}(\Lambda_\text{sp}^{b}) \, , \qquad b > 0 \, .
\end{equation}
The leading term is the first contribution to the mass of a black hole of size $R_\text{S} = \Lambda_\text{sp}^{-1}$. Since the two-derivative effective theory is not wholly dominant at this scale, we do not require that corrections be absent; rather, we require that they do not feature multiplicative logarithms, rather strictly smaller powers of $\Lambda_\text{sp}$. Such terms arise predominantly from local operators in the effective action\footnote{Terms of this type can be argued to dominate in the presence of large extremal black holes \cite{Arfaei:2023hml}.}, but non-local terms arising from fluctuations of massless fields lead to similar corrections \cite{Donoghue:1994dn, Xiao:2021zly}. If these are indeed the dominant corrections, since the black hole mass is determined far away from the source via the ADM procedure one expects power-like deviations from the classical result. The calculation for the string tower seems to be at odds with this result, but the source of this conundrum is clear: strings are not light with respect to the UV cutoff, which is parametrically lower than their strong coupling scale. The criterion in eq. \eqref{eq:species_power_correction} can only be trusted when the tower lies below the UV cutoff, so that the semiclassical expansion for the mass-radius relation is (marginally) reliable. In fact, estimating the minimal black hole entropy within string perturbation theory leads to the crossover scale $M_s^{3-d}$ rather than $\Lambda_\text{sc}^{3-d}$, as in 
eq. \eqref{eq:BH_transition}, and no contradictions arise \cite{Bedroya:2022twb}.

In scenarios where numerous fields solely interact through gravity, one might intuitively expect that as one increases the number of states within the system, gravitational forces will always drive the system to collapse into a black hole. However, it turns out that the formation of a minimal black hole is not guaranteed to occur consistently with eq. \eqref{eq:species_power_correction} or with the entropy $S_\text{sp} \sim N_\text{sp} + \mathcal{O}(\log N_\text{sp})$. The latter relation is expected to arise from the large-area EFT expansion $S \sim \frac{A}{4} + \mathcal{O}(\log A)$, where $N_\text{sp} \propto A$ for large horizon area in Planck units. Importantly, within the EFT regime the leading correction is logarithmic, and its coefficient is computable \cite{Sen:2012dw}. When such inconsistencies do arise, an intuitive picture is that a large number of light species adds complexity to the process of minimal black hole formation, since it also increases the minimal length $L_\text{sp}$ that can be resolved. 

Our primary goal is to unravel some conditions under which a tower exhibits a direct thermodynamic connection with a minimal black hole according to the above criteria. To this end, we will consider a variety of towers of species defined by a mass scaling as given in eq. \eqref{eq:general_mass}, namely
\begin{equation}
m_n = m \, f(n) \, ,
\end{equation}
where $m$ is the typical tower scale. The species scale then relates to the maximum allowed level $N$ according to
\begin{equation}
m_N = \Lambda_\text{sp} = m \, f(N) \, .
\end{equation}
Summing the degeneracies $d_n$ yields the number of species
\begin{equation}
N_\text{sp} = \sum_{n=0}^N d_n = \Lambda_\text{sp}^{2-d}
\end{equation}
as usual. Similarly, the species energy is given by
\begin{equation}
E_\text{sp} = \sum_{n=0}^N d_n \, m_n = m \sum_{n=0}^N d_n \, f(n) \, .
\end{equation}
As discussed in the top-down section \ref{sec:top-down}, for a KK tower the species energy indeed fulfills the condition in eq. \eqref{eq:species_power_correction} with $b=1$.
This lesson will prove important when addressing whether a given tower can be consistent with black hole thermodynamics, i.e. whether a given tower can exhibit a transition with a black hole. In this setting, we can ask whether these two towers are the only ones that afford an interpretation in terms of a transition between black holes and species towers. More concretely, we would like to constrain the set of pairs $(d_n, f(n))$ by the following conditions:
\begin{align}\label{Constraints1}
E_\text{sp} &= \sum_{n=0}^N d_n m_n = \gamma \, \Lambda_\text{sp}^{3-d} + \mathcal{O}(\Lambda_\text{sp}) \, , \\
S_\text{sp} & = \Lambda_\text{sp}^{2-d} + \mathcal{O}\left(\log \Lambda_\text{sp}^{2-d}\right)\, .
\end{align}
These conditions stem from an EFT analysis, which means that a tower that does not satisfy them is either inconsistent. However, given a tower at fixed cut-off $\Lambda_{sp}$, or energy $E_{sp}$, then it is not sufficient to ask that the transition can happen only at a specific energy scale. The right constraint to impose is indeed to extend this transition to every energy level $N$, \emph{i.e.} in terms of distance, at every (parametrically large) distance in some space of vacua we are generically considering. 

Combining the above relations on the energy, we can write the following equation:
\begin{equation}\label{Rec1}
\sum_{n=0}^N d_n f(n) = \gamma \left(\sum_{n=0}^N d_n\right) f(N) \, .
\end{equation}
Now, let us consider moving in moduli space toward the infinite-distance limit, in such a way that the allowed level $N \to N+1$ increases by one. Since the degeneracies and masses are the same, one has
\begin{align}
N'_\text{sp} = \sum_{n=0}^{N+1} d_n = N_\text{sp} + d_{N+1}, \\
m_{N+1} = \Lambda'_\text{sp} = m_0 \, f(N+1).
\end{align}
Due to the same considerations as above, we can write
\begin{equation}\label{Rec2}
\sum_{n=0}^{N+1} d_n f(n) = \gamma \left(\sum_{n=0}^{N+1} d_n\right) f(N+1) \, .
\end{equation}
Subtracting eq. \eqref{Rec1} from eq. \eqref{Rec2}, we obtain the following recursive equation,
\begin{equation}\label{RecursiveEq}
f(N+1) = \frac{\sum_{n=0}^{N} d_n}{\sum_{n=0}^{N+1} d_n - \frac{1}{\gamma} d_{N+1}} f(N) \, .
\end{equation}
From \eqref{Rec1}, it is easy to see that
\begin{equation}
\gamma f(N) =  \frac{\sum_{n=0}^{N} d_n f(n)}{\sum_{n=0}^{N} d_n} < \max_{n}{(f(n))} = f(N) \implies \gamma < 1 \, .
\end{equation}
Together with eq. \eqref{RecursiveEq}, this implies that $f(N+1) > f(N)$, which is indeed our hypothesis on $f$.

As for the entropy, the expected asymptotics is based on the relation $N_\text{sp} = \Lambda_\text{sp}^{2-d} \propto A$, where $A$ is the horizon area in Planck units. Similarly to the approach leading to eq. \eqref{RecursiveEq}, 
imposing eq. \eqref{Constraints1} and using the general entropy formula of eq. \eqref{eq:general_species_entropy}, one can obtain a constraint on the leading behavior of the mass spectrum $f(n)$ given the degeneracy $d_n$. For large $N$, one can equivalently differentiate with respect to the control parameter $m$ or $N$, finding
\begin{equation}
    N_\text{sp} \, \frac{f'(N)}{f(N)} \sim \left(a + \frac{b}{N_\text{sp}}\right) \frac{\partial N_\text{sp}}{\partial N}
\end{equation}
for some constants $a \, , \, b$. Integrating this equation one obtains the large-$n$ behavior
\begin{equation}\label{eq:entropy_bootstrap}
    f(n) \propto \left(\sum_{k=1}^n d_k\right)^a\,,
\end{equation}
up to additive $\mathcal{O}\left(\left[\sum_{k=1}^n d_k\right]^{-1} \right) $ corrections. Here, in order that $f(n)$ be monotonic, one need consider $a>0$. As we have seen, the string tower violates this behavior, which is another indication that it is above the UV cutoff and the correct effective state counting must be modified.

Due to the generality of this problem, here we will consider more families of solutions. We shall divide the discussion according to whether the towers are power-like or not, since this behavior mirrors the KK case examined in section \ref{sec:KK_to_BH}. We will observe that the degeneracy of the tower actually fixes the mass spectrum of the tower itself to some extent. Namely, imposing that these towers of species can form a black hole, we find that the degeneracies are accompanied by specific mass spectra. In turn, whenever these mass spectra are allowed, they appear to conform to the emergent string scenario of \cite{Lee:2018urn, Lee:2019wij, Lee:2019xtm}, as we shall see.

\subsection{Power-like towers}\label{sec:power-like_towers}

In order to streamline the ensuing discussion, we divide it in two parts. In the first part we begin with a warm-up exercise, considering towers with constant degeneracy. We will use the constraints introduced above to find the allowed mass spectra $f(n)$. Then, in the second part we generalize the analysis to power-like degeneracies.

\subsubsection{Case 1: constant degeneracies}\label{sec:constant_deg}

Let us begin by considering a tower with constant degeneracy. In this setup, according to eq. \eqref{RecursiveEq}, the value of the degeneracy does not play a significant role for the mass spectrum. The general recursive equation becomes
\begin{equation}
  f(N+1) =  \frac{N}{N+1 -\frac{1}{\gamma}} f(N) \equiv \frac{N}{N + \alpha} f(N),
\end{equation}
where $\alpha = 1 - \frac{1}{\gamma} < 0$. Solving the recursion, we obtain 
\begin{equation}\label{RecKK}
    f(N+1) = \frac{N!}{\prod_{k=1}^N (k+\alpha)} f(1) \, .
\end{equation}
From the positivity of $f$ in eq. \eqref{RecKK}, it is possible to find a lower bound for $\gamma$, namely $\gamma > \frac{1}{2}$. Later, we will explore how subleading corrections modify this bound. The relation \eqref{RecKK} can be recast as
\begin{equation}\label{RecKK2}
    f(N+1) = \frac{\Gamma(N+1)\Gamma(\alpha+1)}{\Gamma(N+1+\alpha)} f(1) \, .
\end{equation}
At first sight, this relation may appear different from usual KK towers. However, in the large $N$ limit, it simplifies to
\begin{equation}
    f(N+1) = f(1) \Gamma(1+\alpha) (N+1)^{-\alpha}\left(1 + \mathcal{O}\left(\frac{1}{N}\right)\right),
\end{equation}
\emph{i.e.} the mass scaling function at large $N$ is power-like, with power $\frac{1}{\gamma}-1$,
\begin{equation}\label{MassScalingKK}
    f(N) \sim N^{\frac{1-\gamma}{\gamma}} \, .
\end{equation}
It is natural to connect eq. \eqref{MassScalingKK} to a KK tower in $p$ extra dimensions, where $f(n) = n^{\frac{1}{p}}$ using the effective parametrization. This identifies
\begin{equation}
    \gamma = \frac{p}{p+1} \, ,
\end{equation}
which nicely matches the result in eq. \eqref{GeneralEnergyRel}.

We can now include subleading corrections of order $\mathcal{O}(\Lambda_\text{sp})$. Consequently, we impose the constraints
\begin{align}\label{Constraints2}
E_\text{sp} &= \sum_{n=0}^N d_n m_n = \gamma \Lambda_\text{sp}^{3-d} + \omega \Lambda_\text{sp} \, , \\
\Lambda_\text{sp}^{2-d} &\equiv \sum_{n=0}^N d_n, \quad \Lambda_\text{sp} = m_0 f(N) \, .
\end{align}
Applying the same rationale as in the preceding discussion, the resulting recursion takes the form
\begin{equation}
    f(N+1) = \frac{\sum_{n=0}^{N} d_n + \sigma}{\sum_{n=0}^{N+1} d_n - \frac{1}{\gamma} d_{N+1} + \sigma} f(N) \, ,
\end{equation}
where $\sigma \equiv \frac{\omega}{\gamma}$ is so far unconstrained. Solving it one obtains
\begin{equation}\label{sol2}
    f(N+1) = \frac{\Gamma(N + \sigma + 1)\Gamma(\alpha +1)}{\Gamma(N+1+\alpha)\Gamma(\sigma +1)}f(1) \, ,
\end{equation}
where now $\alpha \equiv 1 + \frac{\omega-1}{\gamma}$. In this case, additional constraints arise from the properties of the $\Gamma$ function. Specifically, the domain of the above expression is
\begin{equation}
\begin{cases}
\gamma > \frac{1-\omega}{2} \\
1<\gamma < -\omega  \\

\end{cases} \implies
    \begin{cases}
        \gamma > \frac{1}{2/(1-\omega)} < \frac{1}{2} & \textit{if $\omega>0$}, \\
        \gamma > \frac{1}{2/(1-\omega)} > \frac{1}{2} & \textit{if $-1<\omega<0$}, \\
       0  &\textit{if $\omega<-1$}
    \end{cases}
\end{equation}
Hence, it is impossible for $\gamma$ to assume values less than $\frac{1}{2}$ if $\omega \in (-1, 0)$. As before, when considering the large-$N$ limit, we can express eq. \eqref{sol2} as
\begin{equation}
    f(N) = f(1) \frac{\Gamma(\alpha+1)}{\Gamma(\sigma+1)}\, N^{\sigma - \alpha}(1 + \mathcal{O}\left(1/N\right)) \, .
\end{equation}
This, once again, results in the same mass scaling factor as seen in eq. \eqref{MassScalingKK}, namely
\begin{equation}
    f(N) \sim N^{\frac{1-\gamma}{\gamma}} \, .
\end{equation}
As observed, the introduction of subleading corrections of the order $\mathcal{O}(\Lambda_\text{sp})$ to the species energy alters the parameter space in the setup without affecting the mass scaling factor. Considering general quantum/higher-derivative corrections in a gravitational EFT \cite{Donoghue:1994dn, Xiao:2021zly}, we can infer that the correction has to be positive, $\omega>0$, and
\begin{equation}\label{eq:p_bound}
    \frac{1}{2} \leq \gamma \leq 1 \implies p \geq 1 \, .
\end{equation}

Another way to confirm the same result is using eq. \eqref{eq:general_species_entropy} for constant degeneracy. Imposing that the second term be at most of order $N_\text{sp} = N$, differentiating with respect to $N$ (thought of as a function of the control parameter $m$) one finds 
\begin{equation}
    \frac{f'(N)}{f(N)} \leq \mathcal{O}(1) \, ,
\end{equation}
which is saturated by $f(n) \sim n^\beta$ for a constant $\beta$. Imposing consistency with the species energy, one obtains $\beta = \frac{1-\gamma}{\gamma} = \frac{1}{p}$. These towers behave as KK towers, according to the discussion in section \ref{sec:KK_to_BH}. Let us emphasize that the bound in eq. \eqref{eq:p_bound} pertains to an effective parametrization of KK towers, which includes not only the integers associated to ordinary KK towers but also any rational $p_\text{eff}$ arising from combined KK towers as in \cite{Castellano:2021mmx}. However, towers that formally correspond to less than one compact dimensions are excluded.

\subsubsection{Case 2: power-like degeneracies}\label{sec:power-like_deg}

We now extend the preceding discussion to a power-like degeneracy $d_n = n^\alpha$. The resulting recurrence equation for $f(N)$ takes the form
\begin{equation}
f(N) = \frac{\sum_{n=1}^{N-1} n^\alpha}{\sum_{n=1}^N n^\alpha - \frac{1}{\gamma} N^\alpha} f(N-1) \, .
\end{equation}
In the limit of large $N$, this leads to the solution
\begin{equation}
f(N) = \frac{2^{\alpha + 1}}{N^{\alpha+1}} \frac{\Gamma(2)\Gamma(N+1-b)}{\Gamma(N+1)\Gamma(2-b)} f(2) \, ,
\end{equation}
where $b \equiv \frac{1+\alpha}{\gamma}$. To ensure the positivity of the argument in the $\Gamma$ functions, $\gamma$ must lie below the pole in $f(2)$, which occurs for $1+2^\alpha-\frac{1}{\gamma} \, 2^\alpha = 0$. It follows that
\begin{equation}\label{eq:gamma_bound}
\frac{1}{2} \leq \frac{2^\alpha}{1+2^\alpha} < \gamma < 1 \, .
\end{equation}

Since the degeneracy must be $d_n \geq 1$, this strongly constrains the exponent to $0 \leq \alpha < 1$. We can expand the solution for $f(N)$, finding
\begin{equation}\label{eq:pow-pow_f}
f(N) \sim N^{(1+\alpha)\left(\frac{1-\gamma}{\gamma}\right)} \, .
\end{equation}
This result shows that, in the limit as $\alpha \to 0$, we recover the outcome discussed in the previous section. Additionally, it provides insights into the constraints on the mass scaling, suggesting that exponential degeneracy can, at most, have an exponential mass scaling, as will be examined in the next section.

Finally, for these tower we can calculate the species energy for this tower. Letting the exponent in eq. \eqref{eq:pow-pow_f} be $\beta$, for $f(n) = n^\beta$ one has
\begin{eqnarray}
    E_\text{sp} = \frac{\Lambda_\text{sp}}{N^\beta} \sum_{n=1}^{N} n^{\alpha+\beta} \sim \frac{1+\alpha}{1+\alpha+\beta} \, \Lambda_\text{sp}^{3-d} + \frac{\beta (1+\alpha)^{\frac{\alpha}{1+\alpha}}}{2} \, \Lambda_\text{sp}^{\frac{1-\alpha (d-3)}{1 + \alpha}} \, .
\end{eqnarray}
For this tower, the species scale is
\begin{equation}
    \Lambda_\text{sp} \sim m^{\frac{1}{1+(d-2) \frac{\beta}{1+\alpha}}} \, ,
\end{equation}
which once more identifies $\beta = \frac{1+\alpha}{p}$ and $\gamma = \frac{p}{p+1}$, where $p$ mimicks the number of extra dimensions associated to a KK tower. Using eq. \eqref{eq:gamma_bound}, one finds

\begin{eqnarray}
    p \geq 1 \, ,
\end{eqnarray}

which is indeed bounded below by 1\footnote{Equality can be obtained adding the first correction to the species energy as in the previous case.}. This result once more covers not only (effective parametrizations of) single KK towers, where $p$ is the number of compactified dimensions, but also includes rational effective values $p_\text{eff}$ arising from combined KK towers \cite{Castellano:2021mmx}.

We can also find the expression for the entropy in eq. \eqref{eq:anyenergy_species_entropy}. Expanding in terms of $N_\text{sp}$, the first subleading correction cancels, leaving the second one. In particular, one finds
\begin{equation}
    S_\text{sp} \sim \frac{1+\alpha+\beta}{1+\alpha} \, N_\text{sp} \, ,
\end{equation}
and the resulting entropy is compatible with the EFT expansion for black holes.

All in all, once more we recover KK-like towers by the consistency of minimal black hole thermodynamics. In order to go further, one would have to include subleading corrections, which perhaps would be able to detect the difference between different effective parametrizations such as the one used in eq. \eqref{eq:KK_loop_estimate} or the one in \cite{Castellano:2022bvr}. 
One potentially subtle difference between these is that the domain of summation is not the same as the exact representation in terms of KK quantum numbers, since the effective KK levels would be mapped to (functions of) norms $\abs{\mathbf{k}}$ of lattice vectors $\mathbf{k}$ are in general not integers. Our approach cannot distinguish between these various domains, but on the flip side it can cover any type of KK tower, as shown for instance by the bounds on the effective parameter $p_\text{eff} \geq 1$. As a consistency check, the standard KK parametrization with $\alpha=p-1$ gives $f(n) = n$, \emph{i.e.} $\beta = 1$ as expected.

\subsection{Subpower-like and superpower-like towers}\label{sec:nonpower-like_towers}

In the preceding section we have learned that power-like scalings and degeneracies mimic the KK-like case, including the allowed range of exponents. Thus, in order to further probe how black hole thermodynamics constrains light towers, first we extend the preceding analysis to a power-like mass spectrum with logarithmic degeneracy, and then we consider cases where the mass spectrum $f$ is subpower-like or superpower-like. These examples serve as a useful starting point to understand whether (and how) minimal black hole thermodynamics can constrain towers of species from a bottom-up perspective.

\subsubsection{Case 1: logarithmic degeneracies}\label{sec:log_deg}
We can also consider the case of logarithmic degeneracies $d_n$, keeping a power-like mass spectrum $f(n) = n^b$. This power-like spectrum with logarithmic degeneracy can be excluded by thermodynamic consistency, combined with additional considerations. Indeed, the species scale can be expressed in terms of the mass scale $m$ as
\begin{equation}\label{polLogSC}
    \Lambda_\text{sp} \sim {m^{\frac{1}{1+(d-2)b}}}{\left(\log \frac{1}{m}\right)^{-\frac{b}{1+(d-2)b}}} \, ,
\end{equation}
while the species energy and the entropy behave as
\begin{equation}\label{eq:log-pow_species_energy}
    E_\text{sp} = m \left(\frac{N^{b+1}}{b+1} \, \log N - \, \frac{N^{b+1}}{(b+1)^2} \right) \propto \Lambda_\text{sp}^{3-d} \left( 1 + \mathcal{O}\left( \frac{1}{\log \Lambda_\text{sp}}\right)\right) \, .
\end{equation}
\begin{eqnarray}
    S_\text{sp} = (1+b) \, N_\text{sp} - \, \frac{b^2}{1+b} \, \frac{N_\text{sp}}{\log N_\text{sp}} \, .
\end{eqnarray}
Much like in the case of the logarithmic masses analyzed in section \ref{sec:log_mass}, we interpret this result to mean that either the tower is light with respect to the UV cutoff but inconsistent with the expected EFT expansion of the black hole mass-radius relation, or it is heavier than the UV cutoff. In the latter case, the S-matrix bootstrap arguments we discuss in section \ref{sec:complementary_arguments} imply that this spectrum cannot be consistent. It is the degeneracy that is ruled out by causality, since the tower needs higher-spin excitations to UV-complete gravitational scattering \cite{Camanho:2014apa, Afkhami-Jeddi:2018apj, Arkani-Hamed:2020blm}. Actually, this towers differs from the others in another aspect: the correction in eq. \eqref{eq:log-pow_species_energy} is negative.

An additional interesting case comprises logarithmic degeneracies and mass spectra, with $d_n = \log^a n$ and $f(n) = \log^b n$. In this case, the species scale is given by
\begin{equation}
    \Lambda_\text{sp} \sim m \, \log^b\left(\frac{1}{m}\right) \, ,
\end{equation}
independently of $a$. As we have learned in the preceding cases, this logarithmic behavior in the species scale leads to logarithmic terms in thermodynamical quantities, \emph{e.g.} the species energy
\begin{equation}
    E_\text{sp} \sim \Lambda_\text{sp}^{3-d} \left( 1 + \frac{b}{d-2} \, \frac{1}{\log \Lambda_\text{sp}} \right) .
\end{equation}
As a result, we expect this tower to be either inconsistent, with no possible UV completion, or not light with respect to the UV cutoff of a putative UV completion. As before, the latter case is ruled out by causality arguments, as we shall review in section \ref{sec:complementary_arguments}.

\subsubsection{Case 2: logarithmic spectra}\label{sec:log_mass}

As a simple example of a subpolynomial spectrum, let us consider a logarithmic tower, where $f(n) = \log{(n)}$ and the degeneracy is constant, $d_n = \beta$. Much like the KK case with the effective parametrization of \cite{Castellano:2022bvr}, the number of species is $N_\text{sp} = \beta N$ in terms of the maximum level $N$. The species scale is then related to $m$ and $N$ according to
\begin{equation}\label{logKK}
m_N \equiv \Lambda_\text{sp} = m \log{(N)} \quad\longrightarrow{}\quad m = \frac{\Lambda_\text{sp}}{\log{(N)}} \, .
\end{equation}
The resulting energy associated with these species is then
\begin{equation}
E_\text{sp} = \sum_{n=1}^N \beta \, m \log{(n)} = \beta \, m \log{\Gamma(N+1)} \overset{N \gg 1}{\sim} \beta \, m (N \log N - N) \, .
\end{equation}
Expressing the result in terms of the species scale, we arrive at
\begin{equation}\label{eq:energy_log_tower}
E_\text{sp} = \Lambda_\text{sp}^{3-d} \left( 1 + \frac{1}{d-2} \frac{1}{\log{(\Lambda_\text{sp})}} + \mathcal{O}\left(\frac{1}{\log{\log{\Lambda_\text{sp}}}}\right)\right) \equiv \gamma(\log{(\Lambda_\text{sp})}) L_\text{sp}^{d-3} \, .
\end{equation}
This result is reminiscent of the string tower, and indeed the species scale
\begin{equation}\label{eq:scale_log_tower}
    \Lambda_\text{sp} \sim (d-2) \, m \, \log \frac{1}{m}
\end{equation}
has an analogous expression. Similarly, including the second subleading correction to the species energy, the entropy behaves as
\begin{equation}\label{eq:entropy_log_tower}
    S_\text{sp} \sim N_\text{sp} + \frac{1}{2} \, \frac{N_\text{sp}}{\log^2 N_\text{sp}} \, .
\end{equation}

One could conceive that adding a degeneracy may change the result. However, if the degeneracy $d_n \geq 1$ the species energy cannot decrease. An easy example to see this is a linear degeneracy $d_n = n$. The mass spectrum \eqref{logKK} does not change in this setup, and so we can write
\begin{eqnarray}\label{hyperfactorialKK}
    E_\text{sp} = \beta \, m \sum_{n=1}^{N} n \log{(n)} = \beta m \log{\mathcal{H}(N)} \, ,
\end{eqnarray}
where $\mathcal{H}(N)$ is the hyperfactorial  $\mathcal{H}(N) \equiv \prod_{n=1}^N n^n$.
For large $N$, we can write \eqref{hyperfactorialKK} as
\begin{eqnarray}
    E_\text{sp} \sim \beta \, \frac{m}{2} \left[ \frac{N(N+1)}{2} \log{N} - \frac{N^2}{2} + \frac{1}{12}\right] = \gamma(\log(\Lambda_\text{sp}))
    \, \Lambda_\text{sp}^{3-d} +\mathcal{O}(\Lambda_\text{sp})\, ,
\end{eqnarray}
which indeed features the same type of subleading term. It may be useful to note that implementing the modified counting of states of \cite{Dvali:2009ks} outlined at the end of section \ref{sec:string_to_BH} the spurious terms in between $\Lambda_\text{sp}^{3-d}$ and $\Lambda_\text{sp}$ disappear.
With the same approach it is possible to show that for every degeneracy $d_n = n^a$ and mass $f(n) = \log(n)$ one arrives at the similar result
\begin{eqnarray}
    \Lambda_\text{sp} \sim m \log\left(\frac{1}{m}\right) .
\end{eqnarray}
Analogously to the case studied in the preceding section, the above results suggest that if these logarithmic towers are light with respect to the UV cutoff, they are inconsistent with the expected EFT expansion of the black hole mass-radius relation and entropy, encoded in eqs. \eqref{eq:species_power_correction} and \eqref{eq:entropy_log_tower}. Alternatively, the tower may be heavier than the UV cutoff, which may modify the expected mass-radius relation. This case is similar to the string tower -- in fact, from the point of view of black hole thermodynamics they appear to be essentially indistinguishable. Despite these similarities this particular tower, as well as other inequivalent towers, is likely ruled out by other considerations. Namely, the S-matrix bootstrap arguments which we will discuss in section \ref{sec:complementary_arguments} imply that the case of constant degeneracy is inconsistent with causality of UV-complete perturbative graviton scattering. The case of linear degeneracy may be allowed by this argument, since in this case the degrees of freedom of spin-$j$ fields grow as $j^{d-3} = j = d_j$. In this case, it seems plausible that the correct UV cutoff be given by the mass scale of the tower, $\Lambda_\text{UV} = m$, similarly to the case of the string tower \cite{Afkhami-Jeddi:2018apj}. Indeed, the degeneracy grows exponentially in the mass, although the spacings between levels are different. At any rate, it seems unlikely that the logarithmic spectrum be allowed by unitarity, which strongly constraints the possible spectra \cite{Arkani-Hamed:2020blm, Cheung:2022mkw, Geiser:2022exp}. In particular, the rather general infinite-product family of ansatze of \cite{Geiser:2022exp} turns out to only be consistent when the mass spectrum is the string spectrum. Still, it would be interesting to determine whether this tower can be conclusively excluded by this argument for more general ansatze, or by using a bottom-up modified state counting along the lines of \cite{Dvali:2009ks}, which takes into account that heavier states ought to contribute gradually less to the effective number of excitations that participate in the formation of a black hole. Moreover, even if a set of such towers were compatible in this sense, this can be ascribed to the fact that the thermodynamic quantities are invariant under certain types of effective reparametrizations of the tower \cite{Castellano:2022bvr}, as we shall see in more detail below.

Another related consideration is that the ``emergent'' metric for moduli fields controlling $m$ is given by (see \emph{e.g.} \cite{Castellano:2022bvr, Blumenhagen:2023yws} for the general framework)
\begin{equation}
    ds^2 \sim \frac{dm^2}{m^2} \, ,
\end{equation}
and thus the distance to boundary point $m=0$ diverges logarithmically. In other words, the exponential scaling predicted by the distance conjecture for $m$ holds, while the exponential scaling for $\Lambda_\text{sp}$ does not.

Finally, within towers with logarithmic mass spectra, we consider exponential degeneracies $d_n = e^{\alpha n}$. The species scale
\begin{equation}
    \Lambda_\text{sp} \sim m \, \log \log m^{-1}
\end{equation}
differs qualitatively from the preceding case, and it cannot be a KK tower. Nonetheless, the species energy is still asymptotic to $\Lambda_\text{sp}^{3-d}$, since
\begin{equation}\label{eq:sum_bounds}
    e^{\alpha N} \, \log N \leq \frac{E_\text{sp}}{m} = \sum_{n\leq N} e^{\alpha n} \, \log n \leq \log N \sum_{n \leq N} e^{\alpha n} \sim \frac{e^{\alpha}}{e^\alpha-1} \, \log N \, e^{\alpha N} \, .
\end{equation}
Since the degeneracy is now exponential, the purely bottom-up approach bases on black hole thermodynamics that we employ cannot exclude this tower. However, since the spectrum is logarithmic, it seems very unlikely that such a tower can provide a consistent higher-spin unitarization of gravity, as discussed in section \ref{sec:complementary_arguments}. In particular, fairly general classes of ansatze rule it out explicitly \cite{Caron-Huot:2016icg, Geiser:2022exp, Geiser:2022icl}, although it would be interesting to explore this case further. In this context, the arguments of \cite{Afkhami-Jeddi:2018apj} would suggest that the UV cutoff of a putative completion of this theory be $m$ itself; similar considerations apply to other towers.

\subsubsection{Case 3: exponential spectra}\label{sec:exp_mass}

Next, we consider the simplest example of a superpolynomial mass spectrum, namely an exponential tower characterized by a mass scaling function of the form $f(n) = e^{\alpha n}$ and a constant degeneracy $d_n \equiv \beta$. Now the relation between the tower scale and the species scale reads
\begin{equation}
m_N \equiv \Lambda_\text{sp} = e^{a N } m \qquad \longrightarrow \qquad m = \frac{\Lambda_\text{sp}}{e^{aN}} \, .
\end{equation}
The computation of the species energy in this case leads to
\begin{equation}
E_\text{sp} = \sum_{n=0}^N \beta \, m \, e^{an} = \mathcal{O}(1) \, \Lambda_\text{sp} \, ,
\end{equation}
which readily implies that the species energy tends towards zero as we approach the boundary of the moduli space. In essence, within this scenario, the formation of a black hole becomes unfeasible even in terms of the leading contribution to the energy. Despite an large number of species $N_\text{sp} \sim \frac{1}{\alpha} \log m^{-1} \gg 1$ within the species scale
\begin{equation}\label{eq:species_scale_1-exp}
    \Lambda_\text{sp} \sim \left[\log{\left(\frac{1}{m}  \right)}\right]^{-\frac{1}{d-2}} \, ,
\end{equation}
they cannot compensate the decreasing quantum gravity cutoff as $m \to 0$. Another manifestation of this is the fact that the emergent metric for moduli fields controlling $m$ reads \cite{Castellano:2022bvr, Blumenhagen:2023yws}
\begin{equation}\label{ExpDist}
    ds^2 \sim \Lambda_\text{sp}^{d-2} \, \left(\frac{dm}{m}\right)^2 \, ,
\end{equation}
and thus the distance to $m = 0$ diverges as $\log \log \frac{1}{m}$\footnote{This propriety seems to always hold when the tower cannot form any single-particle-like bound states.}. This means that the approach to masslessness is not exponentially fast in the distance, which violates the distance conjecture\footnote{More precisely, only the exponential scaling is violated. This type of scaling seems to be related to the presence of Einstein gravity at low energies, rather than \emph{e.g.} higher-spin gravity. Relatedly, in AdS settings the holographically dual statement is that the exponential scaling seems to be associated with large-$N$ gauge theory rather than vector models \cite{Basile:2022sda}.}.

From this example we can indeed understand the physics behind the exclusion, but it can be generalized for a polynomial degeneracy $d_n = n^\beta$ and mass $m_n = e^{n^a}$. We momentarily restrict to $a<1$ in order to estimate sums with the Euler-Maclaurin expansion. In this case number of species is
\begin{eqnarray}
    N_\text{sp} = \sum_{n=0}^N n^\beta \equiv H^{(-\beta)}_N \sim \frac{N^{\beta +1}}{\beta + 1} \, , 
\end{eqnarray}
while the species energy is
\begin{eqnarray}
    E_\text{sp} = m \sum_{n=0}^N e^{n^a} n^\beta \sim \frac{1}{a} \, \Gamma\left(\frac{\beta + 1}{a}, -N^a \right),
\end{eqnarray}
where $\Gamma(a, z)$ is the incomplete Gamma function. In the large-$N$ limit we can write
\begin{eqnarray}
    E_\text{sp} \sim \Lambda_\text{sp}^{\frac{(\beta + 1)(3-d) + (d-2)a}{\beta + 1}} \, .
\end{eqnarray}
From the above relation we can see that it is not possible to form a black hole for any value of $\beta$ and $a$. Actually, this conclusion extends also to $a=1$, using an argument similar to the one in eq. \eqref{eq:sum_bounds}. Analogously, it is possible to calculate the emergent distance in eq. \eqref{ExpDist} obtaining the same result. Hence, these cases are indeed excluded by the thermodynamic picture\footnote{It may useful to note that the limit $\beta \xrightarrow{} \infty$ restores the right scaling. This case will be explained in the following sections.}. The same line of reasoning also excludes logarithmic degeneracies $d_n = \log^\beta(n)$.

\subsubsection{Case 4: exponential degeneracies}\label{sec:exp_deg}

Now we explore the mass scaling function for a tower with exponential degeneracy $d_n = e^{an}$. Degeneracies of this type have been observed in the bubbling geometries of \cite{Hatsuda:2023iof}. The recurrence equation for $f(N)$ in this case is
\begin{equation}
f(N) = \frac{\sum_{n=1}^{N-1} e^{an}}{\sum_{n=1}^N e^{an} - \frac{1}{\gamma} e^{aN}} f(N-1) \, .
\end{equation}
In the large $N$ limit, this leads to
\begin{equation}
f(N) \sim \frac{(e^{2a} b -1)}{(be^a)^{N}}  f(2) \, ,
\end{equation}
where $b = 1 - \frac{e^a-1}{e^a\gamma}$. To ensure monotonicity and positivity of $f(N)$, we find the constraint
\begin{equation}\label{eq:exp-exp_positivity}
\frac{e^a}{e^a+1} \leq \gamma < 1 \, .
\end{equation}
Since $a>0$ for the degeneracy to grow, the above inequalities also imply $\gamma \geq \frac{1}{2}$. This domain allows us to derive
\begin{equation}\label{eq:exp-exp_spectrum}
f(N) \sim e^{cN} \, ,
\end{equation}
where $c = -a - \log{b} \geq 0$. Thus, an exponential degeneracy requires an exponential mass spectrum. The species energy for $f(n) \sim e^{cn}$ then reads
\begin{eqnarray}
    E_\text{sp} = \frac{e^c(e^a -1)}{e^{a+c}-1} \Lambda_\text{sp}^{3-d} + \mathcal{O}(\Lambda_\text{sp}) \, .
\end{eqnarray}
Similarly, the entropy evaluates to
\begin{equation}
    S_\text{sp} = \frac{b}{e^a-1} \, N_\text{sp} + \mathcal{O}(\log N_\text{sp}) \, ,
\end{equation}
which is compatible with minimal black hole thermodynamics. Finally the species scale is given by
\begin{equation}
    \Lambda_\text{sp} \sim m^{\frac{1}{1+(d-2)c/a}} \, ,
\end{equation}
which looks like the relation between the $\Lambda_{sp}$ and $m_{KK}$ for $\frac{a}{c} = p \geq 1$, where $p$ is the number of extra dimensions. Another consistency check is that the emergent distance toward the limit $m=0$ is again logarithmically divergent, confirming the exponential decay of $m$ and $\Lambda_\text{sp}$. To wit, the emergent metric is given by
\begin{equation}
    ds^2 \sim \left[\sum_n e^{\alpha n} \, \left( e^{c n}\right)^{d-2} \right]\, m^{d-2} \, \frac{dm^2}{m^2} \sim N_\text{sp}^{1 + (d-2) \frac{c}{\alpha}} \, m^{d-2} \, \frac{dm^2}{m^2} \sim \frac{dm^2}{m^2} \, .
\end{equation}
Similarly, the loop estimates in eq. \eqref{eq:KK_loop_estimate} produce $\Lambda_\text{sp}$ as an upper bound to the UV cutoff, as in the KK case. This tower thus appears to be a KK-like tower in disguise: indeed, the parametrization in eq. \eqref{eq:exp-exp_spectrum} by thermodynamic consistency matches that of \cite{Castellano:2022bvr}. To wit, for suitable functions $F$ for which summation by parts produces subleading terms,
\begin{eqnarray}
    \sum_n d_n \, F(n) = \sum_n e^{\alpha n} F(e^{cn}) \sim N_\text{sp} \, F(N_\text{sp}^{\frac{1}{p}}) \sim \sum_n F(n^{\frac{1}{p}}) \, .
\end{eqnarray}
Let us remark that, to leading order, the agreement between such sums is not surprising, and it follows from the Euler-Maclaurin formula together with integration by parts. Physically, the effective parametrizations are not unique, and thus it is not particularly remarkable that one can pass to a power-like parametrization of this type insofar as the leading contributions are concerned. However, for KK-like towers we found that the subleading terms yield consistent ranges of allowed parameters. Particularly indicative of the KK-like nature of these towers is the agreement of the exponent $\frac{1}{p}$ in the argument of $F$, which appears in the species scale in precisely the same fashion, consistently with the identification of $p$ extra dimensions decompactifying (or, more generally, an effective $p_\text{eff} \geq 1$ as in \cite{Castellano:2021mmx}). These results point to the conclusion that this tower be some sort of effective parametrization of a KK-like tower, but it would certainly be interesting to understand whether bottom-up methods can shed some more light on its character.

The overarching insight gleaned from the analyses presented in this section highlights that the formation of black holes is contingent on the presence of a sufficiently large number of states within the cutoff region. When the number of microstates is substantial, gravitational collapse may be achieved, and the effective description in terms of species towers of the resulting black holes is not unique. However, the thermodynamic picture constrains towers that are light relative to the UV cutoff to be effective reparametrizations of a KK tower.
Conversely, if the gravitating system does not undergo collapse, the number of microstates, delineated by the count of states and their associated energy, displays a volume-dependent behavior: this is the so-called Wheeler's bag of gold seen from another prospective. On the other hand, a logarithmic scaling results in an excessive number of states residing within the cutoff, thus yielding observable corrections that rival the leading-order effects. All in all, these considerations already hint to the conclusion that not all towers are allowed by minimal black hole thermodynamics. As we emphasized, these thermodynamic arguments only exclude towers which are light with respect to the UV cutoff. Among these, the consistent ones exhibit KK-like characteristics in the species scale, energy and entropy. However, when considering towers which are light relative to the Planck scale but not relative to the UV cutoff, one has to also rely on additional arguments. In the next section we present some of them.

\subsubsection{Case 5: generalized string-like spectra}\label{genStringlikeSpectra}

As we have shown in the preceding section, the only way to form a black hole with a light tower of exponential degeneracy is to have a particular exponential mass scaling. This calculation does not exclude towers with $\Lambda_\text{UV} \ll \Lambda_\text{sp}$, in which the effective counting of states should be modified \cite{Dvali:2009ks}. In this section we will consider a generalization of the string spectrum, \emph{i.e.}
\begin{eqnarray}
    d_n = e^{\lambda n^\alpha} \, , \qquad f(n) = n^\beta \, .
\end{eqnarray}
For $\alpha = \frac{p}{p+1}$, degeneracies of this type arise in the semiclassical quantization of $p$-branes \cite{Alvarez:1991qs, KalyanaRama:2003cz, Hayashi:2023txz}. With this setup the number of species is
\begin{eqnarray}\label{numbSpeciesSS}
    N_\text{sp} = \sum_{n=1}^N e^{n^\alpha} \sim \frac{1}{\alpha} \, \Gamma\left(\frac{1}{\alpha}, -N^\alpha \right) + \frac{1}{2} \,  e^{N^\alpha} \, ,
\end{eqnarray}
while the species scale is
\begin{eqnarray}
    \Lambda_\text{sp} = N^\beta \, m \, .
\end{eqnarray}
We restrict to $\alpha < 1$ in order to estimate sums with the Euler-Maclaurin formula as done in the preceding cases. Firstly, our goal is to express the species scale in terms of the mass parameter $m$. In order to do so, in the large-$N$ limit we expand eq. \eqref{numbSpeciesSS} according to
\begin{eqnarray}
    N_\text{sp} \sim e^{N^\alpha}\left( \frac{N^{1-\alpha}}{\alpha} + \frac{1}{2} \right) .
\end{eqnarray}
We can start studying the case $\alpha<1$, such that the second term is subleading. To the same order in the asymptotic expansion, one finds
\begin{eqnarray}
    N^\alpha \propto \log(\frac{1}{m}) +\mathcal{O}\left(\log\log\frac{1}{m}\right),
\end{eqnarray}
analogously to the string case for $\alpha = \frac{1}{2}$. However, in this general case the two exponents may be different. Hence,
\begin{eqnarray}\label{eq: genStringSc}
    \Lambda_\text{sp} \sim m \log^{\frac{\beta}{\alpha}}\left(\frac{1}{m}\right) \, .
\end{eqnarray}
From this result, we can see that species scale obtained in the case of the string tower can be obtained when $\alpha=\beta$, without any further constraints on their individual values.

In order to corroborate the expectation that this tower cannot be parametrically lighter than the UV cutoff, we now compute the asymptotics of the species energy. Once more, we use the Euler-Maclaurin formula combined with
\begin{equation}
    \int^N x^\beta e^{\lambda x^\alpha} \, dx = \frac{N^{1-\alpha+\beta}}{\lambda \alpha} \, e^{\lambda N^\alpha} \left(1 + \frac{1-\alpha-\beta}{\lambda \alpha N^\alpha} + \mathcal{O}(N^{-2\alpha})\right) .
\end{equation}
This time, the leading correction may \emph{a priori} arise from the first-derivative term in the Euler-Maclaurin expansion. Taking this into account, the additional terms take the form
\begin{align}
    \frac{E_\text{sp}}{m} = \sum_{n \leq N} e^{\lambda \, n^\alpha} \, n^\beta & = \frac{N^{1-\alpha+\beta}}{\lambda \alpha} \, e^{\lambda N^\alpha} \left(1 + \frac{1-\alpha-\beta}{\lambda \alpha N^\alpha} + \mathcal{O}(N^{-2\alpha})\right) \\
    & + N^\beta \, e^{\lambda N^\alpha} \left( \frac{1}{2} + \frac{1}{12} \left(\frac{\beta}{N} + \alpha \lambda N^{\alpha-1}\right) \right) + \dots \nonumber
\end{align}
The dots represent additional terms in the Euler-Maclaurin expansion, which we will discuss shortly. From the above expression, one also obtains an expansion for $N_\text{sp}$ setting $\beta = 0$. Expressing the species energy as a function of $N_\text{sp}$ the difference between the two Euler-Maclaurin remainders is proportional to
\begin{equation}
    N^{-\beta} \frac{d^2}{dN^2} N^\beta \, e^{\lambda N^\alpha} - \, \frac{d^2}{dN^2} e^{\lambda N^\alpha} \sim N^{\alpha-2+\beta} \, e^{\lambda N^\alpha} \, . 
\end{equation}
Thus, relative to $N^\beta e^{\lambda N^\alpha}$, the subleading terms in the integral are $\mathcal{O}(N^{1-3\alpha})$, whereas the ones coming from the remainder are $\mathcal{O}(N^{\alpha-2})$. Thus, for $\alpha < \frac{3}{4}$ the former dominates, whereas for $\alpha > \frac{3}{4}$ the latter dominates. All in all,
\begin{align}
    \frac{E_\text{sp}}{\Lambda_\text{sp}} & = N_\text{sp} - \, \frac{N^{1-\alpha}}{\lambda \alpha} \, e^{\lambda N^\alpha} \, \frac{\beta}{\lambda \alpha N^\alpha} + \frac{e^{\lambda N^\alpha}}{12}\, \frac{\beta}{N} + e^{\lambda N^\alpha} \, \mathcal{O}(N^{1-3\alpha}, N^{\alpha-2}) \\
    & = N_\text{sp} + \beta \, \frac{N^{1-\alpha}}{\lambda \alpha} \, e^{\lambda N^\alpha} \left( \frac{\lambda \alpha}{12} \, N^{\alpha-2} - \, \frac{1}{\lambda \alpha} \, N^{-\alpha} + \mathcal{O}(N^{-2\alpha},N^{2\alpha-3}) \right) \, .
\end{align}
For $\alpha < 1$ the $N^{-\alpha}$ term in the parenthesis dominates over all other corrections, and thus one has
\begin{equation}
    E_\text{sp} \sim \Lambda_\text{sp}^{3-d} \left(1 - \, \frac{\beta}{\lambda \alpha} \, N^{-\alpha} \right) \sim  \Lambda_\text{sp}^{3-d} \left( 1 + \, \frac{\beta}{\alpha(d-2)} \, \frac{1}{\log \Lambda_\text{sp}} \right) .
\end{equation}
Therefore, this tower appears to violate the EFT expansion of the mass-radius relation, as expected from the transcendental relation between degeneracy and mass. The arguments reviewed in section \ref{sec:complementary_arguments} are weaker in this case, since the (asymptotic) string spectrum is contained in this class. It would be interesting to explore additional bottom-up arguments to determine whether, among this family, only the string spectrum survives consistently.

\subsection{Complementary arguments for heavy towers}\label{sec:complementary_arguments}

Towers which are not parametrically lighter than the UV cutoff are not amenable to the above arguments, but the analysis of section \ref{sec:typical_HD_scales} and section \ref{sec:KK_to_BH} shows that they cannot be KK towers. Therefore, their inclusion cannot lead to a weakly coupled quantum field theory in higher dimensions. However, we still require that the theory be weakly coupled, based on factorization \cite{Stout:2022phm} supporting the observed patterns in the string landscape. In this setting, using perturbative S-matrix bootstrap methods one can strongly constrain the spectrum of perturbative species, now light with respect to the Planck scale but not with respect to the UV cutoff $\Lambda_\text{UV}$. While this procedure applied to the UV-completion of gauge theories can yield a variety of different spectra\footnote{Furthermore, consistency of multiparticle factorization is even more rigid, ruling out deformations of the Koba-Nielsen amplitude even in the gauge sector \cite{Arkani-Hamed:2023jwn}.} \cite{Cheung:2022mkw, Cheung:2023adk, Cheung:2023uwn}, applying it to gravity leads to constraints that are dramatically more rigid. To begin with, postulating a stringy spectrum $m_n \propto \sqrt{n}$ only allows very sparse and \emph{discrete} deformations from the Virasoro-Shapiro amplitude \cite{Cheung:2023adk, Arkani-Hamed:2020blm} when UV-completing tree-level $2 \to 2$ graviton scattering\footnote{One can argue that either such deformations are inconsistent with higher-point and/or higher-loop scattering or they arise from some asymptotically flat string background \cite{Arkani-Hamed:2020blm}, since string perturbation theory has no non-dynamical free parameters.}.

More importantly for us, however, the spectrum itself is extremely constrained: perturbative unitarity requires introducing an infinite tower of massive higher-spin states to unitarize graviton scattering, as shown in \cite{Camanho:2014apa, Afkhami-Jeddi:2018apj}. In fact, any finite number of additional massive particles generates worse violations of causality, which in \cite{Camanho:2014apa} translates into a bound for a certain kinematic restriction of an amplitude. In this setting, the UV cutoff would then be given parametrically by the mass scale of the tower \cite{Afkhami-Jeddi:2018apj}.

These considerations already rule out subpolynomial degeneracies, since the degrees of freedom of a spin $j \gg 1$ grow as $j^{d-3}$. Furthermore, as a result of including this higher-spin tower, the completed amplitude takes the form
\begin{equation}\label{eq:amplitude_1}
    A_\text{tree}^{2 \to 2} \propto \frac{1}{stu} \, \prod_j \frac{(s+r_j)(t+r_j)(u+r_j)}{(s-m_i^2)(t-m_i^2)(u-m_i^2)} \, ,
\end{equation}
which manifestly implements crossing symmetry, an infinite tower of exchanged particles and the appropriate low-energy limit $A_\text{EFT} \propto \frac{1}{stu}$. As discussed in \cite{Arkani-Hamed:2020blm}, the residue at the pole at, say, $s=m_j^2$ has itself poles at $t=m_i^2$ and $u=m_i^2$. Since $s+t+u=0$, this means that there is also a pole at $t = -m_i^2 -m_j^2$, which must be canceled as well. Thus, the roots $\{r_i\}$ contain all of these combinations \cite{Arkani-Hamed:2020blm}. The simplest way to satisfy these constraints on the numerator is to fix the string spectrum $m_j^2 = M_s^2 \, j$, and it seems much more difficult to deform away from it with respect to the non-gravitational case \cite{Cheung:2022mkw, Cheung:2023adk, Cheung:2023uwn}, although asymptotic stringy Regge trajectories in Yang-Mills scattering are also entailed with certain assumptions \cite{Caron-Huot:2016icg}. A sharper argument was given in \cite{Geiser:2022exp}. Namely, generalizing the infinite-product ansatz in eq. \eqref{eq:amplitude_1} to the form 
\begin{equation}\label{eq:amplitude_2}
    A_\text{tree}^{2 \to 2} = \frac{W(s,t,u)}{s t u} \prod_{n \geq 1} \frac{1 + A_n (st + tu + us) + B_n \, stu}{(1-s/\lambda_n)(1-t/\lambda_n)(1-u/\lambda_n)} \, ,
\end{equation}
with $W = 1 + \mathcal{O}(s,t,u)$ devoid of zeros and poles, one shows that the amplitude is only consistent when $\lambda_n = n$, namely the mass spectrum is the string spectrum. This conclusion is also corroborated by other similar considerations \cite{Geiser:2022icl, Cheung:2022mkw} on the kinematic obstruction to deformations imposed by triple crossing symmetry. For instance, Coon-like $q$-deformations of the Virasoro-Shapiro amplitude are excluded \cite{Geiser:2022icl}. While this does not constitute a complete proof that any possible weakly coupled UV-completion of graviton scattering involves the string spectrum, there seem to be very stringent constraints in this direction.

Putting all these considerations together, towers of species that are parametrically lighter than the Planck scale but heavier than the UV cutoff appear strongly constrained to be stringy. As we have shown in our analysis, towers which are consistent with minimal black hole thermodynamics appear to behave in a KK-like fashion, and thus should be light with respect to the UV cutoff. On the other hand, as discussed in section \ref{sec:string_to_BH} and section \ref{sec:log_mass}, heavy towers require a modified state counting for the UV cutoff relative to the species scale defined by a state counting, which ought to remove multiplicative logarithms. The above considerations suggest that if these towers are consistent they should arise as effective reparametrization of the string tower. However, the constraints on degeneracies and spectra are far more stringent, for instance ruling out the logarithmic degeneracies studied in section \ref{sec:log_deg}.

\section{Conclusions}\label{sec:conclusions}

In this paper we derived a consistency condition on towers of species in quantum gravity from the bottom up, requiring the possibility of a transition/correspondence with minimal black holes. This characterization may not be unique, as we pointed out, but the behavior and scaling of macroscopic quantities such as the species scale exhibit some patterns. Namely, whenever the thermodynamics of the ensemble of light species is consistent with that of a minimal black hole, the towers appear to behave as a KK tower as can be detected from our considerations. The species scale $\Lambda_\text{sp}$ allows a straightforward identification of the number of "extra dimensions", the (loop estimates of) the UV cutoff $\Lambda_\text{UV}$ coincide with $\Lambda_\text{sp}$ and the emergent metric yields an exponential falloff of all relevant scales at infinite distance. 
The correct string limit is also obtained by the formal limit $p \xrightarrow{} \infty$, where $\Lambda_\text{sp} \sim m_\text{tow}$ and the species thermodynamics returns the BH/string transition point, {i.e.}
\begin{eqnarray}
    T_\text{sp} \sim M_s \equiv T_{\text{Hag}}, \quad E_\text{sp} \sim M_s^{3-d} = \frac{M_s}{g_s^2}
\end{eqnarray}
in Planck units. In this limit, the UV cutoff of the EFT is squeezed by the mass gap of the tower and the species scale, in agreement with the bootstrap results of \cite{Camanho:2014apa, Caron-Huot:2016icg, Afkhami-Jeddi:2018apj, Caron-Huot:2022ugt}.

Summarizing, the only two equivalence classes of tower which can thermodynamically correspond to a black hole (or equivalently, for which it is possible to have a BH/species correspondence) are the ones expressed by an effective number of extra dimensions $\hat{p}$ according to
\begin{eqnarray}
    \Lambda_\text{sp} \sim m_{\text{tow}}^{\frac{\hat{p}}{\hat{p}+d-2}}\, , \quad \hat{p} \geq 1
\end{eqnarray}
\begin{eqnarray}
    \Lambda_\text{sp} \sim m_{\text{tow}} \, , \quad \hat{p} = + \infty
\end{eqnarray}
Intriguingly, the constraint $\hat{p} \geq 1$ is automatically obtained by internal consistency. Namely, it arises merely from asking for a physical transition between a generic tower of species and a black hole, without any reference to extra dimensions! This result points to the enticing idea that every consistent tower with a gap below the UV cutoff be nothing but a reparametrization of a KK tower in a UV completion, since our approach is sensitive to its macroscopic features and not to the detailed structure captured by subleading orders. Towers gapped at the cutoff are captured by the formal $\hat{p} \to \infty$ limit, where a sequence of increasingly steep power-like degeneracies can approach an exponential behavior. The other possibilities at weak gravitational coupling (which is definitionally realized in species limits) would be towers of the higher-spin type, which are required by causality in this scenario and are gapped at the cutoff \cite{Camanho:2014apa, Caron-Huot:2016icg, Afkhami-Jeddi:2018apj, Caron-Huot:2022ugt}. Independently from their necessity at weak gravitational coupling (absent any lighter towers, which we argued behave like a KK tower), independently unitarity strongly constrains their spectrum \cite{Caron-Huot:2016icg, Geiser:2022exp, Geiser:2022icl, Cheung:2022mkw, Cheung:2023adk, Cheung:2023uwn} and dynamics \cite{Arkani-Hamed:2020blm, Geiser:2022exp, Geiser:2022icl, Arkani-Hamed:2023jwn} leaving almost no wiggle room to deform away from stringy dynamics. All in all, combining these arguments together provides a coherent picture in support of the dichotomy of limits encoded in the emergent string conjecture.

The results of our analysis based on black hole thermodynamics are summarized in table \ref{tab:1}.

\begin{table}[ht!]
\centering
\begin{tabular}{||c|c|c|c|c||} 
\hline
Mass\textbackslash Degeneracy & Log & Power & Exp & Constant  \\ [0.5ex] 
\hline\hline
Log &  \cellcolor{yellow!50} * & \cellcolor{yellow!50}  & \cellcolor{yellow!50} & \cellcolor{yellow!50} * \\
\hline
Power  &  \cellcolor{yellow!50} * & \cellcolor{green!50} & \cellcolor{yellow!50} & \cellcolor{green!50} \\ 
\hline
Exp & \cellcolor{red!50} & \cellcolor{red!50} & \cellcolor{green!50} & \cellcolor{red!50} \\ 
\hline
\end{tabular}
\caption{A summary of our results based on the bottom-up analysis from minimal black hole thermodynamics. Red indicates that the tower is excluded by thermodynamic consistency, \emph{i.e.} it cannot form a minimal black hole. Green indicates that the tower allows a black hole/species transition, and as discussed in the text all these towers are reparametrizations of KK ones encoding $p \geq 1$ effective extra dimensions (and the correct string limit associated to $p \xrightarrow{} \infty$). Yellow indicates that the bottom-up argument \emph{rules out} the possibility that these towers be parametrically lighter than the (usually unknown) UV cutoff. The multiplicative logarithms that arise in this case indicate that the species scale obtained by state counting is above the UV cutoff. However, as denoted by *, subpolynomial degeneracies are still excluded in these cases, since such towers would need to be of the higher-spin type.}
\label{tab:1}
\end{table}

By the same token, towers that are on the surface inconsistent with minimal black hole thermodynamics are either excluded or behave as string towers in some aspects. For instance, this happens when the mass-radius relation in the EFT is violated by multiplicative logarithmic corrections. For string towers, this seems to be an indication that this logarithm ultimately stems from the fact that the tower mass scale is not parametrically lighter than the UV cutoff. Whether this occurs for other types of towers is difficult to determine in general, since it requires some knowledge about a UV completion (if any exist). Namely, either these towers are light with respect to the cutoff -- in which case they violated the expected structure of black hole thermodynamics -- or they are possibly not light. In the latter case the effective number of species under the species scale (or strong coupling scale) is overcounted, and one has to face the necessity of unitarizing observables outside of the QFT framework. Since the physics remains weakly coupled, it is natural to turn to the numerous remarkable results obtained by the bootstrap program. For the cases that are relevant for us, these results exclude all known examples except for the string spectrum, which crucially features Regge trajectories of higher-spin states. However, recent analyses suggest that their characteristic superpolynomial softness does not significantly constrain the IR physics \cite{Haring:2023zwu}.

In summary, the main lesson that we draw from our analysis is that the consistency of gravitational physics is able to constrain the possible UV completions driven by light towers at infinite distance. Indeed, the extent to which these conditions relate degeneracies and mass spectra, as well as their parameter spaces, is even somewhat surprising, since naively one may have not expected thermodynamics to contain relevant information about microscopic degrees of freedom. Yet, it appears that a combination of the holographic properties of black holes and the minimal threshold encoded in the species scale is able to connect macroscopic properties to some features of light towers of species across the transition regime. Indeed, one can view our results as constraints on which towers allow an interpretation of the species scale in terms of minimal black holes with the appropriate mass-radius relation. These deep physical principles distill a substantial part of our current understanding of quantum gravity, which may perhaps provide a rationale for their surprising (albeit partial) effectiveness in the present context. Another consideration to be made in this respect is that the properties of infinite-distance limits, especially the exponential decay of masses relative to the distance on (pseudo-)moduli spaces, is deeply tied to string theory. A more solid bottom-up grounding in the spirit of \cite{Stout:2022phm} is required to truly understand whether the remarkable structure underlying the emergent string scenario of \cite{Lee:2019wij} is a general feature of quantum gravity or it fundamentally hinges or string theory. A broader understanding of the landscape of consistent EFTs and swampland conditions may also provide further evidence toward the universality of string theory \cite{Montero:2020icj, Bedroya:2021fbu}, which would be another avenue toward answering this question.

Our approach is systematic, and thus can be in principle easily generalized to other types of towers. A more thorough investigation of subleading effects of black hole thermodynamics is also important, in order to determine whether they can provide more information about the possible UV completions, if any. On a more practical note, there are multiple ways to extend the results that we have presented in this paper. 
For instance, aside from classifying families of allowed degeneracies and mass spectra, it could be interesting studying small black holes and their thermodynamics in a supersymmetric setup, in order to infer general properties and perhaps even dualities by wielding more powerful quantitative tools. An attempt along these lines, using swampland arguments, has been carried out in \cite{Bedroya:2023xue}. Another interesting path to study is to understand whether it is possible to reliably count states which lie above the UV cutoff, whenever it differs from the strong coupling scale. It would appear difficult to extend, say, the considerations of \cite{Dvali:2009ks} to settings where no UV completion (if one exists) is known. We expect that such a proper calculation can only be set up using bounds reminiscent of UV/IR mixing and holography, as the black hole argument directly incorporates. Bounds of this type may contain some information about the UV cutoff, allowing a proper effective state counting.

\section*{Acknowledgements}

We thank Alek Bedroya, Ralph Blumenhagen, Niccolò Cribiori, Nicolò Risso, Cumrun Vafa, Timo Weigand and Max Wiesner for insightful discussions. 
We are especially grateful to Alvaro Herraez for the multiple discussions and useful advice. We also thank Alek Bedroya, Rashmish K. Mishra and Max Wiesner for informing us of their related upcoming work. D.L. thanks Harvard University for hospitality during completion of this work.
The work of D.L. is supported by the Origins Excellence Cluster and by the German-Israel-Project (DIP) on Holography and the Swampland.

\appendix
\section{Effective number of KK particles from black holes}\label{app:KK_BH}

In this appendix we briefly review the process of black hole evaporation into KK particles in four extended dimensions \cite{Dvali:2009ks} and the effective counting of KK species in this context. Let us begin recalling that, in a thermal regime, the black hole evaporation rate is given by
\begin{eqnarray}
    \frac{dM_\text{BH}}{dt} \sim - \, T^2 N_{\text{eff}} \, ,
\end{eqnarray}
where the temperature is $T \propto 1/R_\text{S}$. The effective number of species $N_{\text{eff}}$ is given by
\begin{eqnarray}
    N_{\text{eff}} \sim \sum_m e^{-\frac{m}{T}} \int_{\abs{y}<R_\text{S}} \abs{\psi^{(m)}(y)}^2 \, ,
\end{eqnarray}
where $y$ denotes position in the $p$ extra dimensions and $\psi^{(m)}$ are the species wavefunctions restricted within the black hole horizon. Indeed, from a black hole perspective, $N_{\textit{eff}}$ is the effective number of species appearing in eq. \eqref{eq:species_scale_definition}. 

In the simple case of an internal square $p$-torus with volume $\mathcal{V}  (2\pi R)^p$, the KK masses $m_\mathbf{k}$ are given by
\begin{eqnarray}
    m_\mathbf{k} = m_\text{KK} \, \abs{\mathbf{k}} = m_\text{KK} \, \sqrt{\sum_{i=1}^p k_i^2} \, ,
\end{eqnarray}
where the KK quantum numbers $\mathbf{k} = (k_1 \, , \, \dots \, , \, k_p)$ are integers and $m_\text{KK} = \mathcal{V}^{\frac{1}{p}} = R^{-1}$.

For a black hole of size $R_\text{S} \ll R$, the number of thermally available KK states per $(4+p)$-dimensional particle is
\begin{eqnarray}
    N_\text{KK}(m<T) = (TR)^p,
\end{eqnarray}
whereas the effective number of particles is
\begin{eqnarray}
    N_{\text{eff}} = N_{4+p} \, (TR)^p \left(\frac{R_\text{S}}{R}\right)^p = N_{4+p} \, ,
\end{eqnarray}
where the $(R_\text{S}/R)^p$ is the overlap between the KK wavefunctions and the black hole horizon and $N_{4+p}$ denotes the number of species obtained with the ordinary counting. Therefore, the effective state counting reproduces the ordinary result for KK modes.

\bibliography{main}
\bibliographystyle{JHEP}

\end{document}